\documentclass[superscriptaddress, twocolumn, prb]{revtex4-2}

\usepackage{amsmath}
\usepackage{amssymb}
\usepackage{graphicx}
\usepackage{color}
\usepackage{hyperref}
\usepackage{multirow}
\graphicspath{{figs/}}

\begin{document}
	
\title{Non-Abelian geometry, topology, and dynamics of a nonreciprocal Su-Schrieffer-Heeger ladder}

\author{Ziyu Zhou}
\thanks{They contribute equally to this work.}
\affiliation{School of Physics, South China Normal University, Guangzhou 510006, China}
\author{Zhi-Cong Xu}
\thanks{They contribute equally to this work.}
\affiliation{School of Physics, South China Normal University, Guangzhou 510006, China}
\author{Li-Jun Lang}
\email{ljlang@scnu.edu.cn}
\affiliation{School of Physics, South China Normal University, Guangzhou 510006, China}
\affiliation{Guangdong Provincial Key Laboratory of Quantum Engineering and Quantum Materials, South China Normal University, Guangzhou 510006, China}
\affiliation {Guangdong-Hong Kong Joint Laboratory of Quantum Matter, South China Normal University, Guangzhou 510006, China}

\date{\today}

\begin{abstract}

Non-Hermiticity naturally breaks down the adiabaticity and thus leads to non-Abelian behaviors in multi-band systems. 
Here, we study how non-Abelian properties emerge in non-Hermitian systems by considering a multi-band non-Hermitian model --- the nonreciprocal Su-Schrieffer-Heeger (SSH) ladder that is formed by coupling two nonreciprocal SSH chains.
Under periodic boundary conditions, we analytically obtain the exact phase diagrams of the geometry of band structure classified by its complex value and gap type, and of the non-Abelian topology based on a newly defined gauge-invariant winding number under the chiral symmetry.
Under open boundary conditions, we find that the bulk-boundary correspondence survives in the thermodynamic limit but breaks down for finite sizes along with the emergence of critical non-Hermitian skin effects when the inter-leg coupling is weak, where the decaying length $\xi$ of the bulk skin modes varies with the system size $L$, satisfying the scale-free power law $\xi\propto L$. 
Finally, we demonstrate the non-Abelian dynamics of a Bloch state subject to an external constant force in the pseudo-Hermitian symmetric regime in comparison with the non-Hermitian Wilson lines.
Our work may stimulate further interests in nontrivial non-Abelian behaviors in non-Hermitian and open quantum systems.

\end{abstract}

\maketitle

\section{Introduction}

In closed condensed matter systems, the geometry and topology of the band structure of a crystal are essential for understanding the electronic dynamics \cite{Xiao_Niu_2010_RMP}, where the key information comes from the Berry connection and the corresponding Berry phase of a Bloch state's adiabatic evolution in momentum space (i.e., $k$ space) of the crystal \cite{Berry_1984_PRSLA}, especially in single bands. 
The geometry and topology of the subspace involving multiple bands (called the non-Abelian geometry and topology) can be well described by the non-Abelian Berry connection (i.e., Wilczek-Zee connection) and the corresponding Wilson lines \cite{Wilczek_Zee_1984_PRL} if the gaps to the other bands are very large. 
In this case, the inter-band evolution (called the non-Abelian dynamics) in the subspace is no longer adiabatic, although the overall evolution in the subspace preserves the adiabaticity against the subspaces of other bands. 
Non-Abelian properties are common in condensed matter systems, such as in topological insulators \cite{Qi_Zhang_2011_RMP,Hasan_Kane_2010_RMP} and graphenes \cite{CastroNeto_Geim_2009_RMP}, which cannot be properly characterized by single-band quantities due to the existence of band degeneracy in $k$ space.  
The non-Abelian topology and dynamics can be traced by Wilson lines in cold-atom experiments \cite{Li_Schneider_2016_Science,Sugawa_Spielman_2021_npjQI} and have been broadly studied with various theoretic methods \cite{Zhang_Zhou_2017_PRA, Chen_Zhang_2020_PRA,DiLiberto_Palumbo_2020_NC,Jiang_Chan_2022_PRB}.

On the other hand, in open quantum systems the short-time dynamics can be effectively described by non-Hermitian Hamiltonians \cite{Ashida_Ueda_AdvPhys_2020,Breuer_Petruccione_2002_Oxford}. 
The relaxation of the Hamiltonian's Hermiticity brings about many unique phenomena to non-Hermitian systems, such as parity-time-reversal (PT) symmetry breaking \cite{Bender_Boettcher_1998_PRL,Bender_2017_RPP}, non-unitary evolution, exceptional points \cite{Bergholtz_Kunst_2021_RMP}, etc., among which the emergence of the so-called non-Hermitian skin effects (NHSEs) \cite{Yao_Wang_2018_PRL} is one important mechanism for the breakdown and thus reconstruction of the bulk-boundary correspondence (BBC) \cite{Lee_2016_PRL,Leykam_Nori_2017_PRL,Shen_Fu_2018_PRL,Gong_Ueda_2018_PRX,Xiong_2018_JPC,Kunst_Bergholtz_2018,Yokomizo_Murakami_2019_PRL,Jin_Song_2019_PRB,Lee_Thomale_2019_PRB}, deepening our understanding of topology and dynamics in non-Hermitian systems \cite{Ashida_Ueda_AdvPhys_2020}.

Since the intrinsic dissipation in non-Hermitian systems naturally breaks down the adiabaticity, leading to the repopulation at different bands, one may ask a question: What are the non-Abelian properties like in non-Hermitian systems?
In Hermitian systems, one of the simplest models with nontrivial non-Abelian properties is the two-leg Su-Schrieffer-Heeger (SSH) ladder \cite{Zhang_Zhou_2017_PRA} that is formed by coupling two SSH chains as its two legs \cite{Su_Heeger_1979_PRL} and thus has four bands. 
Different varieties of the SSH ladder have been theoretically studied for broad interests \cite{Padavic_Vishveshwara_2018_PRB,Yoshida_Kawakami_2018_PRL,Kurzyna_Kwapinaki_2020_PRB,Nersesyan_2020_PRB,Qin_Wang_2023_PRB,Padhan_Mishra_2024_PRB,Parida_Mishra_2024_PRB,Sabour_Kartashov_2024_OL}. 
Because of the nonsymmorphic symmetry, the band structure of a Hermitian SSH ladder always supports crossing points in $k$ space \cite{Young_Kane_2015_PRL,Fang_Fu_2015_PRB,Shiozaki_Gomi_2015_PRB,MartinezAlvarez_FoaTorres_2018_PRB,Yin_Chen_2018_PRA}, which invalidate the single-band description and require the non-Abelian quantities to describe the topology and dynamics.
Here, we generalize this Hermitian SSH ladder to the non-Hermitian regime that instead couples two nonreciprocal SSH chains as its two legs, dubbed the nonreciprocal SSH ladder.
Apparently, this non-Hermitian ladder is also a four-band model like its Hermitian counterpart but with a generally complex spectrum, and the nonreciprocal legs imply the nontrivial non-Hermitian topology regarding NHSEs. Therefore, the nonreciprocal SSH ladder offers a simple non-Hermitian platform to investigate the non-Abelian properties of non-Hermitian systems.
In addition, the coupling of two nonreciprocal SSH chains with different skin depths will lead to the so-called critical NHSE \cite{Li_Gong_2020_NC,Yokomizo_Murakami_2021_PRB,RafiUlIslam_Jalil_2022_PRR,Liu_Li_2024_PRA}. 
In this nonreciprocal SSH ladder, one may also observe how the non-Abelian properties emerge from the inter-leg coupling.

Although several aspects of non-Abelian properties in non-Hermitian systems have been studied, such as non-Abelian Berry connections in pseudo-Hermitian symmetric phases \cite{Zhu_Palumbo_2021_PRB}, knots of non-Hermitian Bloch bands \cite{Hu_Zhao_2021_PRL}, PT symmetry breaking in non-Hermitian non-Abelian lattice models \cite{Ezawa_2021_PRR}, non-Abelian dynamics in dissipative photonic lattices \cite{Parto_Marandi_2023_NC}, and the gain and loss effect in spin-orbit coupled systems \cite{Xu-Zhu-2022}, and also there are some works regarding different versions of non-Hermitian two-leg ladders \cite{Yang_Song_2018_PRB,Liu_Liu_2019_CPB,Wu_Chen_2022_PRA,Mu_Gong_2022_PRB,Tang_Buljan_2024_APLP,Qi_Liu_2023_PRB,Li_Jiang_2024_NC,Chen_Zhang_2024_PRA}, no detailed analysis is found for the non-Abelian geometry, topology, and dynamics of the nonreciprocal SSH ladder.
Therefore, in this paper after introducing the nonreciprocal SSH ladder in Sec. \ref{sec:Ham}, under periodic boundary conditions (PBCs) we analytically obtain the exact phase diagrams of the geometry of band structure classified by its complex value and gap type in Sec. \ref{sec:geometry}, and of the non-Abelian topology based on a newly-defined gauge-invariant winding number in Sec. \ref{sec:topology}, where the BBC in the thermodynamic limit and the critical NHSE with the scale-free power law for finite sizes under open boundary conditions (OBCs) are also demonstrated. 
In Sec. \ref{sec:dynamics}, we study the non-Abelian dynamics of a Bloch state in the pseudo-Hermitian symmetric regime subject to an external constant force in comparison with the non-Hermitian Wilson lines.
Finally, Sec. \ref{sec:conclusion} gives a conclusion and a short discussion on experiments.

\section{The nonreciprocal SSH ladder}
\label{sec:Ham}
\subsection{Effective non-Hermitian Hamiltonian}

\begin{figure*}[tb]
	\includegraphics[width=1\linewidth]{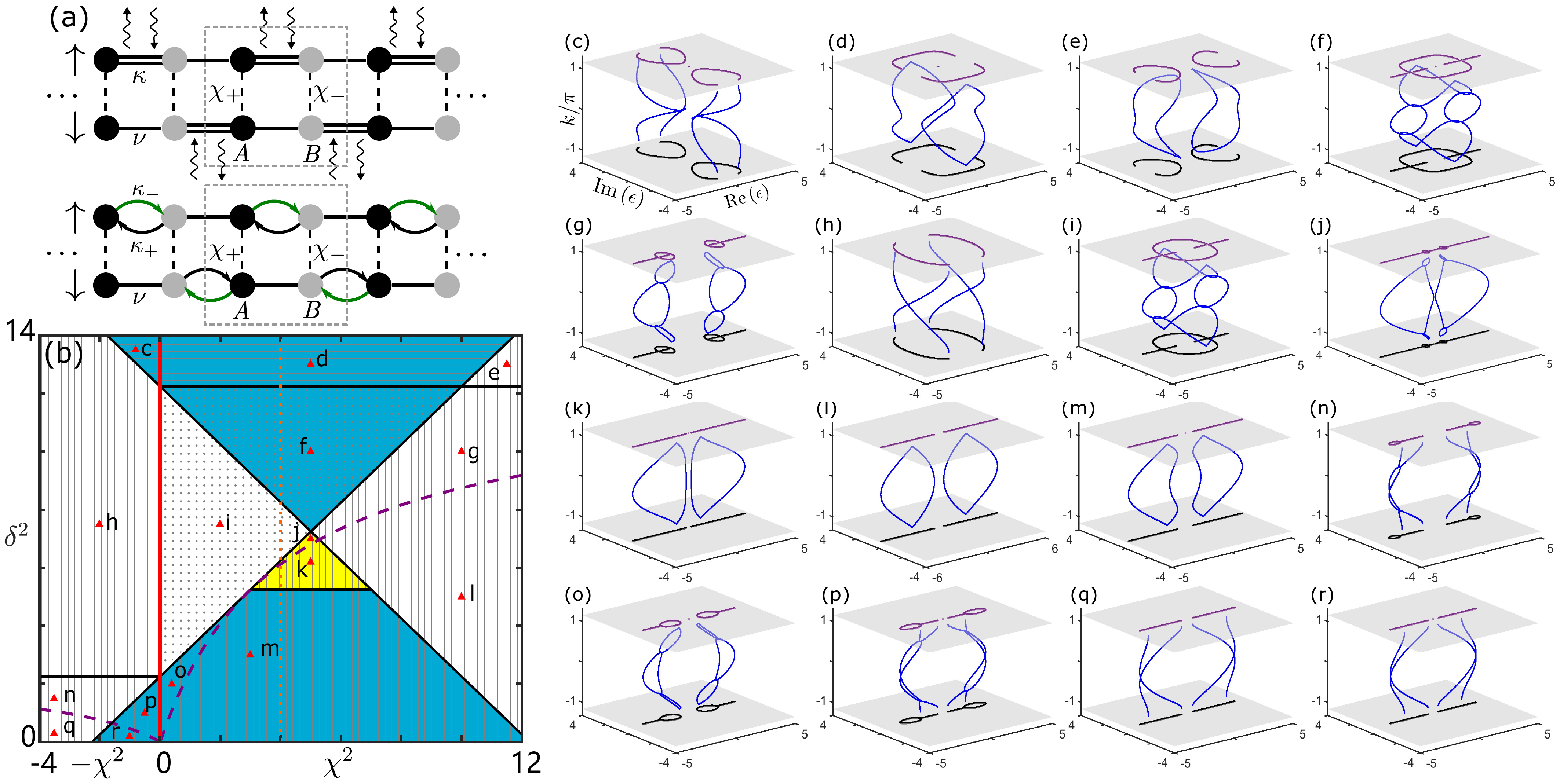}
	\caption{
		(a) Upper panel: Sketch of a dissipative two-leg (denoted by $\uparrow$ and $\downarrow$) SSH ladder coupled to the environment (denoted by the curly arrows), where $\kappa,\nu$, and $\chi_\pm$ label the corresponding hoppings. 
		Lower panel: Sketch of the nonreciprocal SSH ladder represented by $\hat H_\text{NH}$ with $\kappa_\pm$ labeling the nonreciprocal hoppings and other parameters being the same as in the upper panel. 
		The rectangles circle out the unit cells including $A$- and $B$-sublattice sites.
		(b) A typical phase diagram under PBCs for $\kappa=2.5$, composed of the geometry of band structure, characterized by real line gaps (gray vertical lines), imaginary line gaps (gray horizontal lines), and point gaps (gray dots), and of the non-Abelian topology, characterized by the winding number $w=1$ (blue), $-1$ (yellow), and $0$ (white). 
		The black solid lines are the gap-closing positions. 
		Below the purple dashed curves are the pseudo-Hermitian symmetric regimes. 
		The red vertical line at $\chi=0$ separating the negative ($-\chi^2$) and positive $(\chi^2)$ cases are topologically singular in the thermodynamic limit.
		The orange dotted line at $\chi^2=4$ is the parameter path for Fig. \ref{fig2}(a).
		(c)--(r) Typical band structures in the composite 3D space spanned by $k$ and the complex energy plane for parameters $(\chi^2,\delta^2)=(0.8,13.5)$, $(5,13)$, $(11.5,13)$, $(5,10)$, $(10,10)$, $(2,7.5)$, $(2,7.5)$, $(5,7)$, $(5,6.2)$, $(10,5)$, $(3,3)$, $(3.5,1.5)$, $(0.4,2)$, $(0.5,1)$, $(3.5,0.3)$, and $(1,0.2)$, respectively, labeled by red triangles in (b).
		Bottom gray layers: the projection of band structures under PBCs onto the complex energy plane;
		Top gray layers: the corresponding energy spectra under OBCs for $L=200$, where the isolated dots in some gaps are the topological zero-mode end states.
		}
	\label{fig1}
\end{figure*}

Here we couple the system, a two-leg SSH ladder \cite{Zhang_Zhou_2017_PRA} formed by two SSH chains that are relatively shifted by half unit cell, to the environment, as shown in the upper panel of Fig. \ref{fig1}(a). Under the Markov approximation, the dynamics can be well described by the Lindblad master equation \cite{Breuer_Petruccione_2002_Oxford} as follows:
\begin{equation}
	\frac{d\hat \rho}{dt}=-i\big[\hat H, \hat \rho\big] -\sum_{j\sigma}\big(\hat L_{j\sigma}^{\dag} \hat L_{j\sigma} \hat \rho 
	+ \hat \rho \hat L_{j\sigma}^{\dag}\hat L_{j\sigma}
	-2\hat L_{j\sigma} \hat \rho \hat L_{j\sigma}^{\dag}\big),
	\label{eq:Lindblad}
\end{equation}
where $\hat\rho=|\psi(t)\rangle\langle\psi(t)|$ is the density matrix operator of the state  $|\psi(t)\rangle$.
The system Hamiltonian,
\begin{eqnarray}
		\hat{H} =\sum_j\big[&&\kappa\big(\hat{a}_{j \uparrow}^{\dagger} \hat{b}_{j \uparrow}
		+ \hat{b}_{j \downarrow}^{\dag} \hat{a}_{j+1, \downarrow}\big)
		+\nu\big(\hat{b}_{j \uparrow}^{\dag} \hat{a}_{j+1, \uparrow}
		+\hat{a}_{j \downarrow}^{\dag} \hat{b}_{j \downarrow}\big) \notag\\
		&& +\big(\chi_+\hat{a}_{j \uparrow}^{\dag} \hat{a}_{j \downarrow}
		+\chi_-\hat{b}_{j \uparrow}^{\dag} \hat{b}_{j \downarrow}\big)
		+\text {H.c.}\big],
\end{eqnarray}
represents a two-leg SSH ladder with $\hat s^{(\dag)}_{j\sigma}$ ($s=a,b$ and $\sigma =\,\uparrow,\downarrow$) being the annihilation (creation) operator at the $s$-sublattice site of the $j$th unit cell in leg $\sigma$ of the ladder,
and system parameters $\{\kappa,\nu,\chi_\pm\}$, which are set real for simplicity, being the hopping strengths shown in Fig. \ref{fig1}(a). 
The jump operators are set as \cite{Song_Wang_2019_PRL}
\begin{equation}
		\hat L_{j\uparrow}=\sqrt{\delta}\,\big(\hat a_{j \uparrow}+i \hat b_{j \uparrow}\big),  ~
		\hat L_{j\downarrow}=\sqrt{\delta}\,\big(\hat b_{j \downarrow}-i \hat a_{j+1, \downarrow}\big).
		\label{eq:jump_operator}
\end{equation}

For convenience, the master equation (\ref{eq:Lindblad}) can be rewritten as
\begin{equation}
	\frac{d\hat \rho}{dt}=-i\big(\hat H_\text{eff} \hat \rho-\hat  \rho \hat H^\dag_\text{eff} \big) 
	+2\sum_{j,\sigma}\hat L_{j\sigma} \hat \rho \hat L_{j\sigma}^{\dag},
	\label{eq:master_Heff}
\end{equation}
where the summation term is the quantum jump term, and the other term with the effective Hamiltonian, 
$\hat H_\text{eff}\equiv \hat H-i\sum_{j\sigma}{\hat L_{j\sigma}^{\dag}\hat L_{j\sigma}}$, can capture the short-time dynamics before a quantum jump occurs with the aid of post-selection techniques in experiments.
In the explicit form, one can find that the effective Hamiltonian,
\begin{eqnarray}
		\hat{H}_{\mathrm{eff}} 
		&=&\sum_j\Big\{ \kappa_+ (\hat{a}_{j \uparrow}^{\dagger} \hat{b}_{j \uparrow} 
		+\hat{a}_{j+1, \downarrow}^{\dagger} \hat{b}_{j \downarrow})
		+\kappa_- (\hat{b}_{j \uparrow}^{\dagger} \hat{a}_{j \uparrow} 
		\notag\\
		&&+\,\hat{b}_{j \downarrow}^{\dag} \hat{a}_{j+1, \downarrow})
		+\big[\nu (\hat{b}_{j \uparrow}^{\dagger} \hat{a}_{j+1, \uparrow}
		+\hat{a}_{j \downarrow}^{\dagger} \hat{b}_{j \downarrow})
		 +\chi_+\hat{a}_{j \uparrow}^{\dag} \hat{a}_{j \downarrow}
		 \notag\\
		 &&+\,\chi_-\hat{b}_{j \uparrow}^{\dag} \hat{b}_{j \downarrow}
		 +\text {H.c.}\big]\Big\}
		 -i\delta\sum_{j\sigma} (\hat a^\dag_{j\sigma}\hat a_{j\sigma}
		 +\hat b^\dag_{j\sigma}\hat b_{j\sigma}) 
		 \notag\\
		 &\equiv&\hat{H}_\text{NH}-i\delta\hat N,
		 \label{eq:effective Ham}
\end{eqnarray}
is just a non-Hermitian generalization of the two-leg SSH ladder by introducing nonreciprocal hoppings denoted by $\kappa_\pm=\kappa\pm\delta$ and an overall loss related to the total particle number operator $\hat N$. 
Note that in this model, the nonreciprocal hoppings along two legs are engineered to be relatively reversed such that reversed NHSEs occur in different legs when the inter-leg coupling vanishes. 

Since the loss term only contributes an overall decaying factor to the dynamics, without loss of generality, in the following we instead focus on the non-Hermitian Hamiltonian $\hat{H}_\text{NH}$ of the nonreciprocal SSH ladder that is formed by coupling two nonreciprocal SSH chains, as shown in the lower panel of Fig. \ref{fig1}(a).

\subsection{Hamiltonian in $k$ space and the symmetries}
\label{sec:symmetry}

To obtain the non-Abelian properties in $k$ space, 
we transform the non-Hermitian Hamiltonian $\hat H_{\rm NH}$ in Eq. \eqref{eq:effective Ham} 
from real space to $k$ space under PBCs: 
\begin{equation}
	\hat H_{\rm NH}\equiv\hat{\psi}^\dag H \hat{\psi} =\sum_{k\in \text{BZ}}\hat{\psi}_k^\dag H_k \hat{\psi}_k,
	\label{eq:Hk}
\end{equation}
via the discrete Fourier transform, 
\begin{eqnarray}
	\hat{s}_{j\sigma}=\frac{1}{\sqrt{L}}\sum_{k\in\text{BZ}} e^{ikjd} \hat{s}_{k\sigma},
	\label{eq:dFT}
\end{eqnarray}
where $L$ is the number of unit cells, $d$ is the length of the unit cell, and the summation is done over one Brillouin zone (BZ) with $k=2\pi n/L$ ($n\in$ Integer).
The bases in both spaces, respectively, read as
$\hat{\psi}^\dag\equiv(\{\hat a^\dag_{j\uparrow}~\hat a^\dag_{j\downarrow}~\hat b^\dag_{j\uparrow}~\hat b^\dag_{j\downarrow}\})$ and
$\hat{\psi}_k^\dag=\big(\hat{a}^\dag_{k \uparrow}~\hat{a}^\dag_{k \downarrow}~\hat{b}^\dag_{k \uparrow}~\hat{b}^\dag_{k \downarrow}\big)$,
and the Hamiltonian matrix in $k$ space
\begin{eqnarray}
	H_k&=&~
	\left( \begin{matrix}
		\chi_+&		0\\
		0&		\chi_-
	\end{matrix} \right)
	\otimes\sigma_x \notag\\
	&&~+\tau_+\otimes
	\left( \begin{matrix}
		\kappa_+ +\nu e^{-ikd}&		0\\
		0&		\nu+\kappa_+ e^{-ikd}
	\end{matrix} \right) 
	\notag\\
	&&~+\tau_-\otimes
\left( \begin{matrix}
	\kappa_- +\nu e^{ik}&		0\\
	0&		\nu+\kappa_- e^{ikd}
\end{matrix} \right),
	\label{eqh_matrix}
\end{eqnarray}
where $\tau_{x,y,z}$ and $\sigma_{x,y,z}$ are Pauli matrices acting on the sublattice ($a,b$) and the leg ($\uparrow,\downarrow$) spaces, respectively; 
$\tau_0$ and $\sigma_0$ are corresponding identity matrices, and $\tau_\pm\equiv (\tau_x\pm i\tau_y)/2$.
In the following, we set $d=\nu=1$ as the units of length and energy, respectively.

The symmetries are useful tools for analyzing non-Abelian properties.
Due to the underlying symmetries of $\hat H_\text{NH}$ in real space, $H_k$ satisfies the following relations inherited from the corresponding symmetries: 
(a) time-reversal symmetry: $H_k^*=H_{-k}$, due to the reality of all parameters in $\hat H_\text{NH}$;
(b) chiral symmetry: $CH_k C^\dagger=-H_k$
with $C\equiv\tau _z\otimes \sigma _z=C^\dag=C^{-1}$, due to the bipartition of the ladder;
(c) inversion symmetry: $I_kH_kI_k^\dagger=H_{-k}$ with
$I_k=	\left(\begin{array}{cc}
		1 & 0\\
		0 & e^{-ik}
	\end{array}\right)
	\otimes \sigma _x$
satisfying $I_k^\dag=I_k^{-1}$.
These relations ensure that the band structure of $H_k$ has a $D_{2h}$ point group symmetry in the composite three-dimensional (3D) space spanned by three twofold rotating axes: the $k$ axis, the real and the imaginary axes of the complex energy plane. This implies that the gap closing (indicating a phase transition) can only occur at the real or imaginary axes, as shown in Figs. \ref{fig1}(c)--\ref{fig1}(r).
The details of symmetry analysis can be referred to in Appendix \ref{appendix:eta operation}.

Additionally, $H_k$ also has an $\eta_k$-pseudo-Hermitian symmetry (the recap of pseudo-Hermiticity can be found in Appendix \ref{asec:recap}) \cite{Ashida_Ueda_AdvPhys_2020}:
\begin{equation}
	\eta_k H_k\eta_k^{-1}=H_k^\dagger,
	\label{eq:pseudo-Hermitian_k}
\end{equation}
where the invertible Hermitian matrix $\eta_k$ can be explicitly expressed for $\chi_+=\chi_-$ as
\begin{equation}
	\eta_k\equiv 
\left(\begin{array}{cc}
	0 & e^{-ik/2}\\
	e^{ik/2} & 0
\end{array}\right)
\otimes \sigma _x
=\eta_k^{-1}=\eta_k^\dag,
\label{eq:eta_k}
\end{equation}
but it cannot be easily cast in a simple form for $\chi_+\ne\chi_-$.
This $\eta_k$-pseudo-Hermiticity of $H_k$ guarantees that the eigenvalues are either real or appear in complex-conjugate pairs, respectively defining the pseudo-Hermitian symmetric phase and the broken phase, according to whether or not the pair of the right- and left-column eigenvectors $u^{(r/l)}_k$ of $H_k$ satisfies the relation $\eta_k u^{(r)}_k =\lambda_k u^{(l)}_k$ or equivalently $u^{(r)\dag}_k\eta_k u^{(r)}_k=\lambda_k$ with $\lambda_k$ being a nonzero real number.
It is worth noting that even for $\chi_+=\chi_-$, where $\eta_k$ is explicitly written in a simple form in Eq. \eqref{eq:eta_k}, the corresponding pseudo-Hermiticity operation in real space is highly complex, which can be referred to in Appendix \ref{appendix:eta operation}.

Based on these underlying symmetries, we analyze the following non-Abelian properties, including the geometry of band structure with $D_{2h}$ symmetry, the topology under the chiral symmetry, and the dynamics in the pseudo-Hermitian symmetric regime.

\section{Geometry of band structure}
\label{sec:geometry}

The geometry of band structure can be obtained by analytically diagonalizing $H_k$, yielding four energy bands:
\begin{equation}
	\epsilon_k =\pm \sqrt{\alpha_k \pm \sqrt{\beta_k}},
	\label{eq:energy}
\end{equation}
where two ``$\pm$"s are uncorrelated, and the two quantities
\begin{eqnarray}
	\alpha_k&=&\kappa^2+2\kappa \cos k+1+(\chi_+^2+\chi_-^2)/2-\delta^2, \notag\\
	\beta_k&=&4\delta^2(\cos^2 k-1) +2\cos k \big[\kappa\, (\chi_+^2+\chi_-^2) \notag\\
	&&~+(\kappa^2-\delta^2+1)\,\chi_+\chi_-\big] +(\chi_+^2-\chi_-^2)^2/4 \notag\\
	&&~+(\kappa^2-\delta^2+1)(\chi_+^2+\chi_-^2)+4\kappa\, \chi_+\chi_-
	\label{eq:quantities}
\end{eqnarray}
are both real. 
Given the nested form of double-square-root in Eq. \eqref{eq:energy} with real $\alpha_k$ and $\beta_k$, the geometry of band structure can be classified by its complex value (i.e., real, purely imaginary, and complex bands)
and gap type (i.e., real line, imaginary line, and point gaps).
The corresponding conditions are shown in Table \ref{table}.
\begin{table}[h]
	\caption{Classification of the geometry of band structure.}
	\begin{tabular}{ lc}
		\hline\hline
		\multicolumn{2}{c}{Complex value $(\forall~ k)$} \\
		\hline
		Real & $\beta_k\ge 0~\text{and}~\alpha_k\ge\sqrt{\beta_k}$   \\
		Purely imaginary & $\beta_k\ge 0~\text{and}~\alpha_k<-\sqrt{\beta_k}$ \\
		Complex & Others\\
		\\
		\multicolumn{2}{c}{Gap type $(\forall ~k)$} \\
		\hline
		\multirow{2}{*}{Real line} & $(\beta_k\ge 0~\text{and}~\alpha_k>\sqrt{\beta_k})$ \\
		& or $(\beta_k<0~\text{and}~\alpha_k\ne0)$ \\
		Imaginary line & ~$(\beta_k\ge 0~\text{and}~\alpha_k<-\sqrt{\beta_k})$ or $\beta_k<0$~ \\
		Point & Others \\
		\hline
		\hline
	\end{tabular}
	\label{table}
\end{table}
For simplicity but without loss of the typical geometry of band structure, we focus on two symmetric cases of $|\chi_+|=|\chi_-|\equiv \chi>0$, where we call $\chi_+\chi_-=\pm \chi^2$ the positive and the negative cases, respectively. Typical band structures are shown in Figs. \ref{fig1}(c)--\ref{fig1}(r).
Thus, the quantities in Eq. \eqref{eq:quantities} are reduced to
\begin{eqnarray}
	\alpha_k&=&\kappa^2+2\kappa \cos k+1+\chi^2-\delta^2, \notag\\
	\beta_k^{(\pm)}&=&2(1\pm \cos k)\big[\chi^2  (\kappa\pm 1)^2-\delta^2(\chi^2\mp2\cos k+ 2)\big],\notag\\
\end{eqnarray}
where the ``$\pm$'' in $\beta_k^{(\pm)}$ corresponds to the positive and the negative cases, respectively.

Equation \eqref{eq:energy} indicates that the touching of any two bands in $k$ space must satisfy one of the following two conditions: 
(I) $\exists~k\in$ BZ, $\beta_k=0$, 
which indicates an always-existing gapless point $k= \pi \,(0)$ for the positive (negative) case,
and an inequality
\begin{eqnarray}
	\frac{\chi^2(\kappa\pm 1)^2}{\chi^2+4}\le\delta^2\le(\kappa\pm1)^2,
	\label{eq:gapless_I}
\end{eqnarray}
where the `$\pm$' corresponds to the positive and the negative cases, respectively;
(II) $\exists~k\in$ BZ, $\beta_k=\alpha_k^2$,
which leads to the gapless equations:
\begin{equation}
	\delta^2=\left\{
	\begin{aligned}
		&(\kappa-1)^2+\chi^2, \\
		&(\kappa+1)^2-\chi^2,~ \chi^2\le (\kappa+ 1)^2, \\
		&\,\kappa^2-1, ~\kappa\ge1~\text{and}~2(\kappa-1)\le \chi^2\le 2(\kappa+1)
	\end{aligned}
	\right.
	\label{eq:gapless_symmetric}
\end{equation}
for the positive case (i.e., $\chi_+\chi_-=\chi^2$), and
\begin{equation}
	\delta^2=\left\{
	\begin{aligned}
		&(\kappa+1)^2+\chi^2, \\
		&(\kappa-1)^2-\chi^2, ~\chi^2\le (\kappa- 1)^2, \\
		&\,\kappa^2-1, ~\kappa\le-1~\text{and}~2(-\kappa-1)\le \chi^2\le 2(-\kappa+1)
	\end{aligned}
	\right.
	\label{eq:gapless_antisymmetric}
\end{equation}
for the negative case (i.e., $\chi_+\chi_-=-\chi^2$).
These gapless equations along with the upper and the lower bounds in Eq. \eqref{eq:gapless_I} can divide a phase diagram into different geometric regimes of band structure, as shown in Fig. \ref{fig1}(b), which can alternatively be derived from the conditions in Table \ref{table} (no purely imaginary band structure for the positive and the negative cases in this model).
Notably, the lower bound of the inequality \eqref{eq:gapless_I} is just the boundary of the region of real band structures (i.e., the pseudo-Hermitian symmetric phase), which satisfies
\begin{equation}
	\delta^2\le\frac{\chi^2(\kappa\pm1)^2}{\chi^2+4}.
	\label{eq:pseudo-Hermiticity}
\end{equation}
In this region, one can numerically verify for the positive case the condition that $u^{(r)\dag}_k \eta_k  u^{(r)}_k$ is a nonzero real number for each $k$. For the negative case, although a simple expression of $\eta_k $ is lacking, one may verify this condition by resorting to the more complex expression of $\eta_k $ in Appendix \ref{asec:recap}.

\section{Non-Abelian topology under the chiral symmetry}
\label{sec:topology}
\subsection{Winding number under PBCs}

To explore the non-Abelian topology (i.e., the topology involving a subspace of multiple bands), given that $\hat{H}_\text{NH}$ respects the chiral symmetry, we define a gauge-invariant winding number under PBCs (which is essentially different from the commonly used gauge-dependent ones, see Appendix \ref{asec:winding_number} for detailed comparison):
\begin{equation}
	w=\frac{1}{4\pi i}\int_{k \in \text{BZ}} \left[\text{d}\ln\det h_k^{(+)}-\text{d}\ln\det h_k^{(-)}\right],
	\label{eq:winding_main}
\end{equation}
where 
\begin{eqnarray}
	h_k^{(\pm)}	&=&
	\left( \begin{matrix}
		\kappa_\pm+\mathrm{e}^{\mp\mathrm{i}k}&		\chi_\pm \\
		\chi_\mp &		1+\kappa_\mp\mathrm{e}^{\pm\mathrm{i}k}\\
	\end{matrix} \right)
\end{eqnarray}
are defined in the following off-diagonal block Hamiltonian matrix:
\begin{eqnarray}
	H_k^{(b)}&=&
		\left( \begin{matrix}
		0&		h_k^{(+)}\\
		h_k^{(-)}&0
	\end{matrix} \right),
	\label{matblock}
\end{eqnarray}
obtained from the Hamiltonian matrix (\ref{eqh_matrix}) by changing the basis to $(\hat{a}^\dag_{k \uparrow}~\hat{b}^\dag_{k \downarrow}~\hat{b}^\dag_{k \uparrow}~\hat{a}^\dag_{k \downarrow})$.
Define a function $d_k \equiv \det h_k^{(+)}=(2\kappa-\chi_+ \chi_-)+(\kappa^2-\delta^2+1) \cos k+i(\kappa^2-\delta^2-1)\sin k$, which generally forms an ellipse in the complex plane by scanning $k$, the polar angle with respect to the ellipse center, from $0$ to $2\pi$. 
Noting that $\det h_k^{(-)}=d_{-k}$, Eq. \eqref{eq:winding_main} becomes
\begin{equation}
	w=\frac{1}{2\pi i}\int_{k \in \text{BZ}} \text{d}\ln d_k,
\end{equation}
which depends on $\kappa,\,\delta^2$, and $\chi_+\chi_-$.
From the geometrical meaning of the winding number $w$, the topological phase with $w\ne 0$ must satisfy the following condition:
\begin{equation}
	|2\kappa-\chi_+ \chi_-|<|\kappa^2-\delta^2+1| ~~\text{and}~~\kappa^2-\delta^2\ne 1,
	\label{eq:condition2}
\end{equation}
and the topological phase boundaries are determined by
\begin{eqnarray}
	\delta^2&=&(\kappa\pm1)^2\mp \chi_+\chi_-,\notag\\
	\text{and}~~~ 
	\delta^2&=&\kappa^2-1~ (\text{within}~ w\ne 0~ \text{regions}). 
	\label{eq:boundary}
\end{eqnarray}
which, for the positive and the negative cases, are identical to the gapless equations \eqref{eq:gapless_symmetric} and \eqref{eq:gapless_antisymmetric}, indicating that the topological phase transitions must be accompanied by the gap closing.
Since $w$ does not change if one simultaneously inverses both signs of $\kappa$ and $\chi_+\chi_-$, one can only plot the topological phase diagram for $\kappa\ge 0$ with respect to $\chi_+\chi_-$, as an example of Fig. \ref{fig1}(b).

\subsection{BBC and critical NHSE}
\label{sec:critical}

\begin{figure*}[tb] 	
	\includegraphics[width=0.8\linewidth]{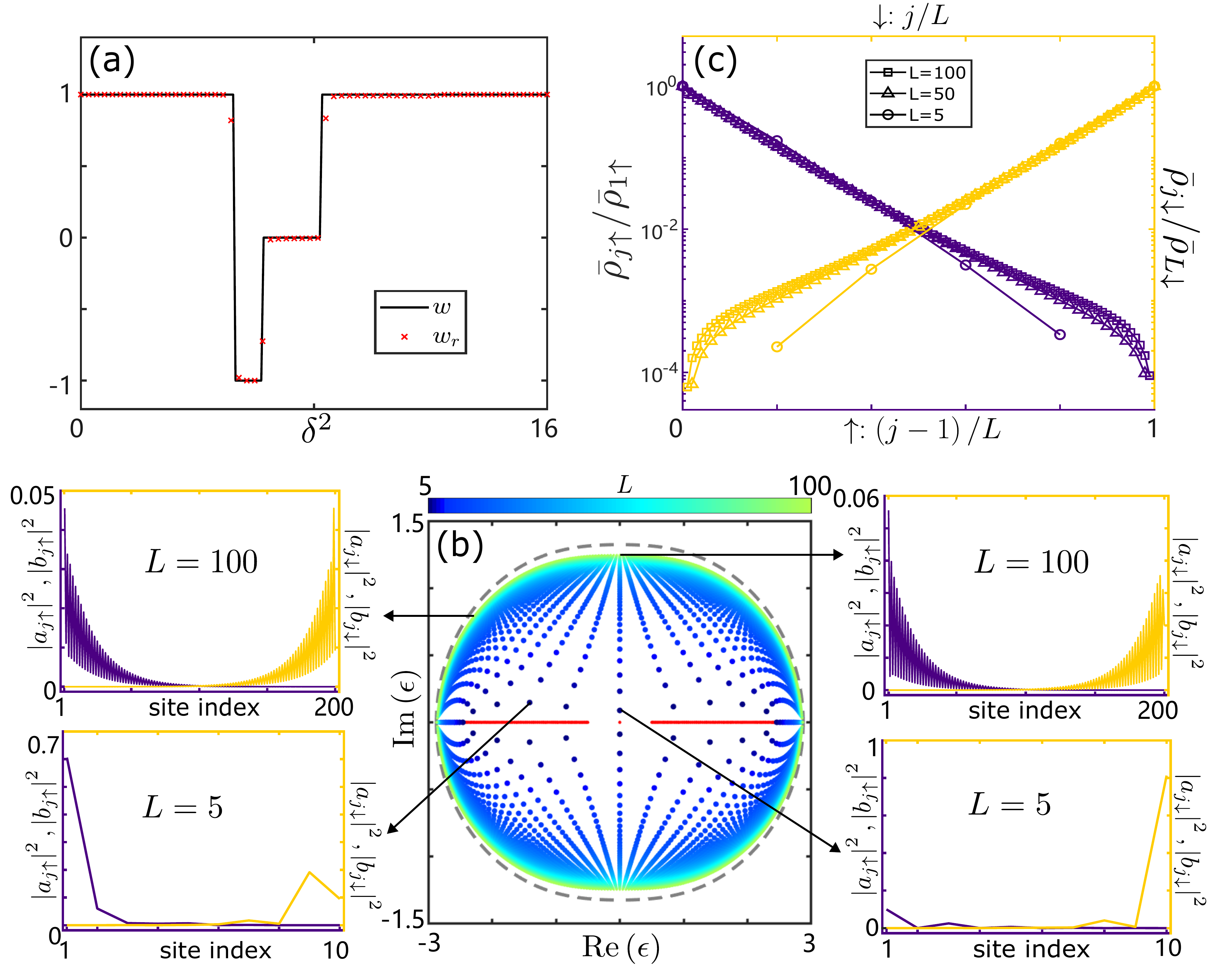}
	\caption{
		(a) Comparison between the winding number $w$ (black solid line) under PBCs and the open-bulk winding number $w_r$ (red crosses) under OBCs along the orange dotted line at $\chi = 2$ in Fig. \ref{fig1}(b), where the parameters for calculating $w_r$ are $(L, L', l ) = ( 200, 160, 20 )$. 
		(b) Energy spectra of $\hat{H}_\text{NH}$ in the complex energy plane under OBCs for $L=5,6,\dots,100$ (dots with colors from dark blue to light green) at weak inter-leg coupling $\chi=0.01$ of the positive case.
		The spectrum (two red solid lines with dots in-between) under OBCs without inter-leg coupling for $L=100$ and the spectrum (gray dashed circle) under PBCs at $\chi=0.01$ of the positive case are also plotted for reference.  
		Insets: Typical density distributions in leg $\uparrow$ (purple) and leg $\downarrow$ (yellow) for the eigenstates indicated by the arrows.  
		The other parameters are $(\kappa, \delta) = (2.5, 2)$. 
		(c) The averaged density distributions versus the scaled unit-cell index for $ L = 5$ (circles), $50$ (triangles), and $100$ (squares) in leg $\uparrow$ (purple) and leg $\downarrow$ (yellow).
	}
	\label{fig2}
\end{figure*}	

To investigate whether the nonreciprocal SSH ladder satisfies the BBC, we numerically plot the spectra under OBCs with $L=400$ unit cells [top layers of Figs. \ref{fig1}(c)--\ref{fig1}(r)], and find that except for the topological zero-mode end states, which are protected by the chiral symmetry, the bulk spectra are identical to those under PBCs [bottom layers in Figs. \ref{fig1}(c)--\ref{fig1}(r)], indicating the survival of BBC in this ladder.
In addition, we also numerically calculate the open-bulk winding number in real space under OBCs \cite{Song_Wang_2019_PRL,Xu-Zhu-2022}:
\begin{equation}
	w_r=\frac{1}{8 L^{\prime}} \operatorname{tr}^{\prime}\hat{C} \hat Q[\hat Q, \hat X],
	\label{Winding_OBC}
\end{equation}
where $\hat C$: $\hat a_{j\uparrow}^{(\dag)}\rightarrow \hat a_{j\uparrow}^{(\dag)}$, $\hat a_{j\downarrow}^{(\dag)}\rightarrow -\hat a_{j\downarrow}^{(\dag)}$,
	$\hat b_{j\uparrow}^{(\dag)}\rightarrow -\hat b_{j\uparrow}^{(\dag)}$, $\hat b_{j\downarrow}^{(\dag)}\rightarrow \hat b_{j\downarrow}^{(\dag)}$ is the chiral operator, 
$\hat Q\equiv\sum_{n}\big[|\phi_n^{(r)}\rangle\langle \phi_n^{(l)}|-\hat C| \phi_n^{(r)}\rangle\langle\phi_n^{(l)}|\hat C \big]$ is the ``$Q$-matrix'' operator constructed by the $n$th chiral-operation paired right/left eigenstates $\big\{| \phi_n^{(r/l)}\rangle,\hat C|\phi_n^{(r/l)}\rangle\big\}$ of $\hat H_\text{NH}$ under OBCs,
and the primed trace means that the trace is done over a central $L'=L-2l$ unit cells out of the whole $L$ unit cells of the ladder with $l$ unit cells being cut off from each end of the ladder to avoid the boundary effects when the numerical calculation is executed.
It is shown in Fig. \ref{fig2}(a) that $w_r$ is close to $w$ for large systems, which hints that there is no BBC breaking and thus no skin effect in the thermodynamic limit $L\rightarrow\infty$.
This can also be reflected by the winding number of energy under PBCs \cite{Gong_Ueda_2018_PRX,Jiang_Chen_2019_PRB}:
\begin{equation}
	w_e=\frac{1}{2\pi i}\int_{k \in \text{BZ}} \text{d}\ln \det (H_k -\epsilon_r I),
	\label{winding_energy}
\end{equation}
where $\epsilon_r$ is an arbitrary reference point in the complex energy plane and $I$ is the identity matrix. 
In our model, that $w_e=0$ is always satisfied due to the inversion symmetry that enforces the energy spectrum to be symmetric with respect to $k=0$ plane, leading to the relation $\det (H_{-k}-\epsilon_r I) =\det (H_k-\epsilon_r I)$.

However, when the inter-leg coupling vanishes (i.e., $\chi=0$), the ladder is decoupled into two independent nonreciprocal SSH chains, which are well known for the breakdown of BBC and the emergence of NHSEs under OBCs.
This indicates that in the thermodynamic limit, the BBC is abruptly restored with the introduction of an infinitesimal inter-leg coupling, which is called the critical NHSE \cite{Li_Gong_2020_NC}.
The critical NHSE leads to a noticeable alteration of the energy spectra and eigenstates under OBCs in the weak inter-leg coupling limit as the system size increases, as shown in Fig. \ref{fig2}(b). 
When the size is small, the energy spectra and eigenstates are close to those of the decoupled nonreciprocal SSH chains under OBCs, i.e., the spectra are linelike and the eigenstates demonstrate NHSEs, indicating that the coupling effects are negligible; 
when the size becomes larger, they converge to those under PBCs, i.e., the linelike spectra expand to loops and the eigenstates tend to be delocalized without NHSEs as Bloch waves, indicating that the weak coupling becomes stronger in the thermodynamic limit. 
Quantitatively, for a finite-size, weakly inter-leg coupled ladder under OBCs, the distribution of bulk eigenstates remains the exponential form $|\phi_j |^2\propto  e^{-|j-j_b|/\xi}$ of the skin bulk states of the nonreciprocal SSH chain, localized in the boundary unit cell $j_b$, but the decaying length $\xi$ of the amplitude $\phi_j$ is size dependent and satisfies the scale-free power law \cite{Li_Gong_2020_NC,Yokomizo_Murakami_2021_PRB}:  $\xi(L) \propto L^c$, unlike the nonreciprocal SSH chain, where $\xi^{-1} =2^{-1}\ln |\kappa _+/\kappa _-|$ is size independent \cite{Yao_Wang_2018_PRL}. 
To show the power law, we define the averaged density distributions in leg $\sigma$ ($\sigma=\uparrow$ and $\downarrow$) for certain system size with unit-cell number $L$: 
\begin{equation}
	\bar\rho_{j\sigma}\equiv \frac{1}{L}\sum_{n=1}^{L}\big(|a_{j\sigma n}|^2+|b_{j\sigma n}|^2\big),
\end{equation}
where the amplitudes $a_{j\sigma n}$ and $b_{j\sigma n}$ are defined in the eigenvalue equation $H\phi_n^{(r)}=\epsilon_n\phi_n^{(r)}$ with the $n$th right-column eigenvector $\phi_n^{(r)}=(\{a_{j\uparrow n},a_{j\downarrow n},b_{j\uparrow n},b_{j\downarrow n}\})^T$.
Figure \ref{fig2}(c) shows the averaged density distributions $(\bar\rho_{j\uparrow}/\bar\rho_{1\uparrow}, \bar\rho_{j\downarrow}/\bar\rho_{L\downarrow})$ normalized by the corresponding localized boundary values versus the scaled unit-cell index $j/L$ for different system sizes. 
The collapse into the same lines for, respectively, upper and lower legs verifies the scale-free power law with the power $c=1$.

In addition, the critical NHSEs also have impact on the topological end states.
Since the zero-mode end state protected by the chiral symmetry appears under OBCs when $\delta^2<$($>)$ $\kappa^2-1$ for the nonreciprocal SSH chain with the (non)reciprocal unit cells being the boundary cells \cite{Yao_Wang_2018_PRL}, in the decoupled case the two relatively shifted nonreciprocal SSH chains as a whole always possess end states under OBCs. 
This means that a topological end state [e.g., the bottom right inset of Fig. \ref{fig2}(b)] in small-sized ladder with weak inter-leg couplings under OBCs may evolve into a bulk state [e.g., the top right inset of Fig. \ref{fig2}(b)] of a topologically trivial phase in the thermodynamic limit, of which, according to the phase diagram in Fig. \ref{fig1}(b), the winding number $w=0$, i.e., the BBC is restored in the thermodynamic limit.
At this moment, one may realize that the red vertical line at $\chi^2=0$ of the phase diagram in Fig. \ref{fig1}(b) is singular in the thermodynamic limit.

It is apparent that if we add a leg-dependent potential, the NHSE under OBCs along with the breakdown of BBC will return in the thermodynamic limit, with NHSEs depending on the eigenenergies, and thus the so-called bipolar NHSE naturally appears \cite{Song_Wang_2019_PRL,Qin_Lee_2023_PRB}. 

\section{Non-Abelian dynamics in the pseudo-Hermitian symmetric regime}
\label{sec:dynamics}

\begin{figure*}[tb] 	
	\includegraphics[width=0.8\linewidth]{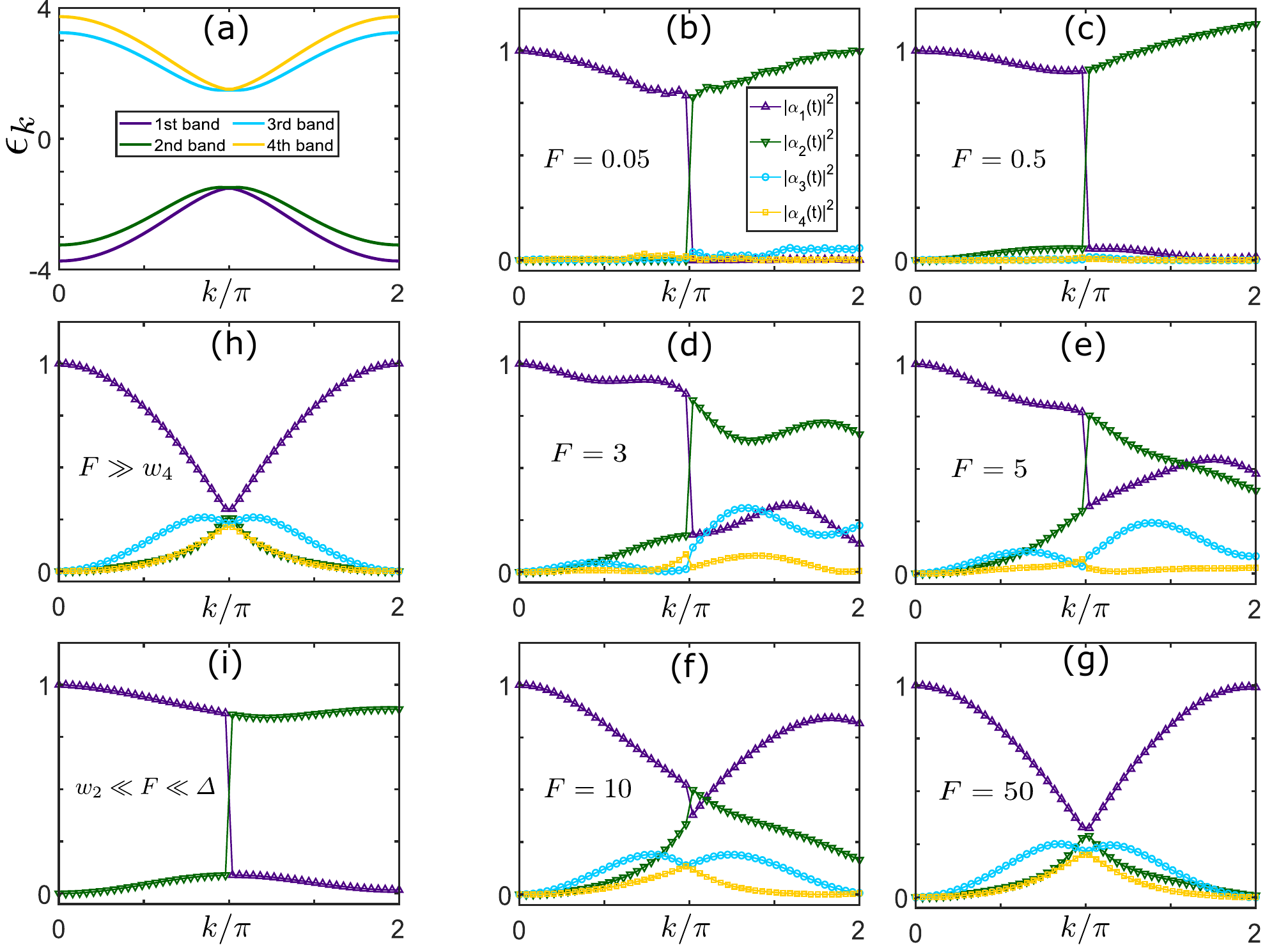}
	\caption{
		(a) A typical real band structure with parameters $(\kappa, \chi^2,\delta^2)=(2.5,0.06,0.06)$ in the pseudo-Hermitian symmetric phase of the positive case. The characteristic quantities  are $(w_2,\Delta,w_4)\approx(0.49,2.97,7.47)$.
		(b)--(g) The evolution of $|\alpha_n(t)|^2$ in $k$ space with an initial state being located in the lowest band at $k=0$ in (a) for various strengths of the force $F=0.05$, $0.5$, $3$, $5$, $10$, and $50$, calculated by Eq. \eqref{eq:solution}.
		(h),(i) The evolution of $|\alpha_n(t)|^2$ in $k$ space with the same initial state as in (b)--(g), but calculated by the four-band and two-band Wilson lines $W_{0\rightarrow k}$ and $W_{0\rightarrow k}^{(-)}$ in Eqs. \eqref{eq:wilson4} and \eqref{eq:wilson2}, respectively. 
		All the evolutions are numerically calculated by dividing the path $k:0\rightarrow 2\pi$ into $49$ equal pieces, and the details of the method can be found in Appendix \ref{asec:wilson_line}.
		}
	\label{fig3}
\end{figure*}	

Since there always exists a band-touching point at $k=\pi\,(0)$ for the positive (negative) case, the adiabatic evolution for a single band cannot work any more in $k$ space. 
Here we focus on the non-Abelian dynamics (i.e., the dynamics involving multiple bands) in the pseudo-Hermitian symmetric regime where all bands are real and thus the dynamics is expected stable without amplitudes decaying or amplifying in the long-time limit, while in the cases with complex bands, the eigenstates with eigenenergies possessing the largest imaginary parts will dominate in the long-time limit.

To investigate the non-Abelian dynamics in $k$ space, we add an external constant force $F$ along the ladder, i.e., the Hamiltonian in Eq. \eqref{eq:Lindblad} becomes $\hat H-F\hat X$, 
where for simplicity the position operator $\hat X$ is defined only with respect to the positions of unit cells, i.e.,
\begin{eqnarray}
	\hat X=\sum_{j\sigma} j(\hat a_{j\sigma}^{\dag}\hat a_{j\sigma}+\hat b_{j\sigma}^{\dag}\hat b_{j\sigma}).
\end{eqnarray}
Following the similar derivation from Eq. \eqref{eq:master_Heff} to \eqref{eq:effective Ham}, the short-time dynamics before a quantum jump occurs can be well captured by the time-dependent non-Hermitian Schr\"odinger equation ($\hbar=1$):
\begin{equation}
	i{\partial_t} |\psi (t)\rangle = (\hat H_\text{NH}-F \hat X)|\psi (t)\rangle,
	\label{eq:schrodinger}
\end{equation}
where $\partial_t\equiv \partial/\partial t$, and the loss term $-i\delta\hat N$ in Eq. \eqref{eq:effective Ham}, which only contributes an overall decaying factor to the dynamics, is ignored.
The initial state is set as a superposition of Bloch states of different bands at $k_0$, and presumably evolves to another superposition at $k(t) = k_0 + Ft$ at time $t$ \cite{Li_Schneider_2016_Science}, i.e., $|\psi(t)\rangle \approx \sum_{n=1}^4 \alpha_n(t) |u_{k(t),n}^{(r)}\rangle$,
where $\alpha_n(t)$ is the amplitude of right Bloch state $|u_{k(t)n}^{(r)}\rangle$ of $\hat{H}_\text{NH}$ in the $n$th band at $k(t)$; we normalize the amplitude at $t=0$, i.e., $\sum_n|\alpha_n(0)|^2=1$.
Thus, the solution to Eq. \eqref{eq:schrodinger} reads as
\begin{equation}
	\alpha(t)= \mathcal{P}\exp\Big\{-i\int_{k_0}^{k} dk\big(\Lambda_{k}/F+A_{k}\big)\Big\}\alpha(0),
	\label{eq:solution}
\end{equation}
where $\alpha(t)=[\alpha_1(t),\alpha_2(t),\alpha_3(t),\alpha_4(t)]^T$, 
$\mathcal{P}$ is the path-ordered operator,
and
\begin{equation}
	A_{k(t)}=-iU_k^{-1}\partial_k U_k\big|_{k=k(t)}
\end{equation}
with $\partial_k\equiv\partial/\partial_k$ is the non-Hermitian non-Abelian Berry connection for the whole bands with the matrices $U_k$ and $\Lambda_k$ being defined in the eigenvalue decomposition
$U^{-1}_k H_k U_k=\Lambda_k\equiv\text{diag}(\epsilon_{k1},\epsilon_{k2},\epsilon_{k3},\epsilon_{k4})$.
Note that the normalization of $\alpha(t)$ at $t\ne 0$ does not hold in general due to the non-Hermiticity of $\hat H_\text{NH}$ and thus the non-unitary evolution of Eq. $\eqref{eq:solution}$. 
The detailed derivation of Eq. \eqref{eq:solution} and the numerical method to solve it can be referred to in Appendix \ref{asec:wilson_line}.
Figures \ref{fig3}(b) to \ref{fig3}(g) show the non-Abelian dynamics in $k$ space under different strengths of force for the initial state being located in the lowest band at $k=0$. The dynamics for other initial states can be referred to in Appendix \ref{seca:initial_states}.

Equation \eqref{eq:solution} shows that the dynamics is closely related to the Berry connection $A_k$.
When $ F\gg w_4$, where $w_4$ is the maximum energy difference between the highest and the lowest bands (i.e., the total band width), $A_k$ dominates the evolution and all four bands can be regarded as being degenerate in the whole BZ. 
Thus, Eq. \eqref{eq:solution} reduces to
\begin{eqnarray}
	\alpha(t)&\approx & \mathcal{P}\exp\Big[-i\int_{k_0}^{k} dk A_{k}\Big]\alpha(0) 
	\notag\\
	&\equiv&  W_{k_0\rightarrow k}\alpha(0)
	=U_k^{-1}U_{k_0}\alpha(0),
	\label{eq:wilson4}
\end{eqnarray}
where 
\begin{eqnarray}
	W_{k_0\rightarrow k}\equiv \mathcal{P}\exp\Big[-i\int_{k_0}^{k} dk A_{k}\Big]
\end{eqnarray} 
is a four-band Wilson line for non-Hermitian systems along the path from $k_0$ to $k$ in $k$ space \cite{Xu-Zhu-2022}.
Equation \eqref{eq:wilson4} means that the evolution only depends on the start and the final points, regardless of the path's selection, as shown in Fig. \ref{fig3}(h), which reproduces the evolution in Fig. \ref{fig3}(g) by Eq. \eqref{eq:solution} under a very strong force.
Note that after one BZ cycle, the state will totally return to the initial state, which can be directly proved by noting that four-band Wilson line in Eq. \eqref{eq:wilson4} becomes identity.

When $ w_2\ll F\ll \Delta$, where $w_2$ is the maximum energy difference between the lowest (highest) two bands and $\Delta$ is the minimum energy difference (i.e., the gap) between the middle two bands, the lowest (highest) two bands can be approximated as being degenerate in the whole BZ and the excitation to the other two bands can be ignored, that is, 
the dynamics can be considered as an adiabatic evolution in the subspace of the lowest (highest) two degenerate bands.
Thus, Eq. \eqref{eq:solution} can be approximately decoupled into two independent evolutions in each subspace: 
\begin{eqnarray}
	\alpha^{(\pm)}(t)&\approx&  \mathcal{P}\exp\Big[-i\int_{k_0}^{k} dk A^{(\pm)}_{k}\Big]\alpha^{(\pm)}(0) \notag
	\\
	&\equiv&  W^{(\pm)}_{k_0\rightarrow k}\alpha^{(\pm)}(0),
	\label{eq:wilson2}
\end{eqnarray}
where the $+~(-)$ labels the upper (lower) $2\times 2$ diagonal block for the labeled matrices or the upper (lower) $2\times 1$ block for the labeled column vectors.
In this case, the non-Abelian dynamics in corresponding subspaces can be well described by two-band Wilson lines for non-Hermitian systems \cite{Xu-Zhu-2022}: 
\begin{eqnarray}
	W^{(\pm)}_{k_0\rightarrow k}\equiv \mathcal{P}\exp\Big[-i\int_{k_0}^{k} dk A^{(\pm)}_{k}\Big],
\end{eqnarray}  
as shown in Fig. \ref{fig3}(i), which roughly captures the qualitative feature of the direct dynamics in Fig. \ref{fig3}(c) calculated by Eq. \eqref{eq:solution}. The quantitative difference between Figs. \ref{fig3}(c) and \ref{fig3}(i) mainly comes from the dissatisfaction of the lower-bound condition $F\gg w_2$.
Different from the four-band Wilson line, the state evolved by the two-band Wilson line in Eq. \eqref{eq:wilson2} does not return to its initial state after one BZ cycle due to the crossing point in the spectrum.
The details of the numerical method for calculating the two Wilson lines can be referred to in Appendix \ref{asec:wilson_line}. 

The non-Abelian dynamics for other strengths of the force in Fig. \ref{fig3} cannot be well captured by the two Wilson lines. 
However, the common feature is that the population in each subspace smoothly transfers from one band to the other across the degenerate point, manifested by the sudden change of the auxiliary lines in Figs. \ref{fig3}(b)--\ref{fig3}(i).
The similar phenomenon also occurs in the non-Abelian dynamics of Hermitian multi-band systems \cite{Zhang_Zhou_2017_PRA}.

\section{Conclusion and discussion}
\label{sec:conclusion}

In conclusion, we investigate the non-Abelian properties of the nonreciprocal SSH ladder that is formed by coupling two nonreciprocal SSH chains. 
Under PBCs, we exactly obtain the phase diagram of the geometry of band structure, including the real and complex bands, as well as the real line, imaginary line, and point gaps, 
and also obtain the phase diagram of the non-Abelian topology under chiral symmetry, characterized by a newly defined gauge-invariant winding number, which is different from the common-used ones. 
Under OBCs, the BBC is found survived in the thermodynamic limit, but breaks down for finite sizes when the inter-leg coupling is weak, where the critical NHSE emerges, obeying the scale-free power law $\xi(L)\propto L$. 
The non-Abelian dynamics of a Bloch state subject to an external constant force in the pseudo-Hermitian symmetric regime, where all bands are real, is also studied in comparison with the two-band and four-band Wilson lines.

In experiments, this nonreciprocal SSH ladder can be potentially realized using a time-multiplexed photonic resonator network as in Ref. \cite{Parto_Marandi_2023_NC} instead by coupling two SSH chains and replacing the jump operators therein by Eq. \eqref{eq:jump_operator}. Alternatively, one may also simulate this ladder using ultracold atoms with the technique of nonreciprocal transport in momentum space \cite{Liang_Yan_2022_PRL,Gou_Yan_2020_PRL}. 
The classical electric circuit is also a good candidate for this ladder \cite{Jiang_Chen_2019_PRB,Xu_Lang_2023_APS}.

\begin{acknowledgments}
We thank L. He for the helpful discussion on the explicit expression of the psudo-Hermitian operator in real space. This work was supported by the National Key Research and Development Program of China (Grant No.~2022YFA1405304), the Guangdong Basic and Applied Basic Research Foundation (Grant No.~2024A1515010188), the National Natural Science Foundation of China (Grant No.~11904109), and the Startup Fund of South China Normal University.
\end{acknowledgments}

\section*{data availability}
The data that support the findings of this article are not publicly available. 
The data are available from the authors upon reasonable request.

\appendix

\section{Recap of the pseudo-Hermitian symmetry}
\label{asec:recap}

To be self-consistent, we recap the concept of pseudo-Hermitian symmetry in this section.
Here, we assume that $\hat H$ is a diagonalizable linear operator acting on the Hilbert space of dimension $N$, which has a set of complete biorthonormal right and left eigenvectors $\{| \phi^{(r/l)}_{n}\rangle\}$ and a discrete spectrum $\{\epsilon_n\}$:
\begin{eqnarray}
		\langle \phi^{(l)}_{n}| \phi^{(r)}_{m}\rangle &=&\delta_{nm},~~
		\sum_{n=1}^{N} |\phi^{(r)}_{n}\rangle \langle \phi^{(l)}_{n}|=\hat I, \notag\\
		\hat H&=&\sum_{n=1}^{N} \epsilon_n |\phi^{(r)}_{n}\rangle \langle \phi^{(l)}_{n}|.
		\label{eq:spectral}
\end{eqnarray}
$\hat H$ is said to be $\eta$-{\it pseudo-Hermitian} \cite{Ashida_Ueda_AdvPhys_2020} if it satisfies the relation:
\begin{eqnarray}
	\hat H^\dag &=&\hat\eta \hat H\hat\eta^{-1},
	\label{eq:pseudo-Hermitian}
\end{eqnarray}
where $\hat \eta$ is a Hermitian invertible operator.
If there exists one $\hat \eta$ such that $\hat H$ is $\eta$-pseudo-Hermitian, $\hat H$ is said to be {\it pseudo-Hermitian} \cite{Ashida_Ueda_AdvPhys_2020}.

This relation implies that the eigenvalues of $\hat H$ are either real or appear in complex conjugate pairs.
It can be proved as follows:
\begin{eqnarray}
	\hat H^\dag \hat \eta|\phi_n^{(r)}\rangle
	=\hat\eta\hat H|\phi_n^{(r)}\rangle
	=\epsilon_n\hat\eta|\phi_n^{(r)}\rangle,
\end{eqnarray}
which implies that $\hat \eta|\phi_n^{(r)}\rangle$ is also a left eigenvector of $\hat H$ with the eigenvalue $\epsilon_n^*$.
Furthermore, the identity
\begin{eqnarray}
	0&=&\langle \phi_n^{(r)}|\hat H^\dag \hat \eta|\phi_n^{(r)}\rangle-\langle \phi^{(r)}|\hat H^\dag \hat \eta|\phi_n^{(r)}\rangle \notag\\
	&=&	\langle \phi_n^{(r)}|\hat\eta|\phi_n^{(r)}\rangle(\epsilon_n-\epsilon_n^*),
\end{eqnarray}
shows that the eigenvalues are either real  ($\epsilon_n^*=\epsilon_n$) or appear in complex conjugate pairs $(\epsilon_n\neq\epsilon_n^*)$ with $\langle \phi_n^{(r)}|\hat\eta|\phi_n^{(r)}\rangle=0$.
For eigenvectors $|\phi_n^{(r)}\rangle$ with real eigenvalues, if there is no degeneracy, it implies that $\langle \phi_n^{(r)}|\hat\eta|\phi_n^{(r)}\rangle\ne 0$, in which case the eigenvectors are called $\eta$-pseudo-Hermitian symmetric states; 
while for eigenvectors $|\phi_n^{(r)}\rangle$ with non-real eigenvalues, it implies that $\langle \phi_n^{(r)}|\hat\eta|\phi_n^{(r)}\rangle= 0$, in which case the eigenvectors are called $\eta$-pseudo-Hermitian broken states. 
For the degenerate eigenvectors with real eigenvalues, we can choose any superposition $|\phi\rangle$ to get any value of $\langle \phi|\hat\eta|\phi\rangle$.
That the spectrum becomes non-real from real accompanied by the change of $\langle \phi_n^{(r)}|\hat\eta|\phi_n^{(r)}\rangle$ from nonzero to zero is called the pseudo-Hermitian symmetry breaking.

Inversely, any Hamiltonian with either real or complex conjugate paired eigenvalues are pseudo-Hermitian. According to this, Eq. \eqref{eq:spectral} can be generally divided into two parts:
\begin{eqnarray}
	\hat H&=&\sum_{\text{Im}\, \epsilon_{n}=0} \epsilon_{n}|\phi^{(r)}_{n}\rangle\langle \phi^{(l)}_{n}| \notag\\
	&&+\sum_{\text{Im}\, \epsilon_{n}>0}\big(\epsilon_{n}|\phi^{(r)}_{n}\rangle\langle  \phi^{(l)}_{n}|+\epsilon_{n}^*| \tilde \phi^{(r)}_{n}\rangle\langle \tilde \phi^{(l)}_{ n}|\big),
	\label{eq:specH}
\end{eqnarray}
where $|\tilde \phi^{(r/l)}_{ n}\rangle $ is the counterpart eigenvectors to $|\phi^{(r/l)}_{ n}\rangle$.
A set of $\hat \eta$'s can be easily expressed using the eigenstates \cite{Mostafazadeh_2002a_JMP}:
\begin{eqnarray}
		\hat\eta&\equiv&\sum_{\text{Im}\, \epsilon_{n}=0} \sigma_{n}|\phi^{(l)}_{n}\rangle\langle \phi^{(l)}_{n}| \notag\\
		&&
		+\sum_{\text{Im}\, \epsilon_{n}>0}\big(|\tilde \phi^{(l)}_{n}\rangle\langle \phi^{(l)}_{n}|+|\phi^{(l)}_{n}\rangle\langle \tilde \phi^{(l)}_{n}|\big),
		\notag\\
		\hat\eta^{-1}&\equiv&\sum_{\text{Im}\, \epsilon_{n}=0} \sigma_{n}|\phi^{(r)}_{n}\rangle\langle \phi^{(r)}_{n}| \notag\\
		&&
		+\sum_{\text{Im}\, \epsilon_{n}>0}\big(|\phi^{(r)}_{n} \rangle \langle \tilde \phi^{(r)}_{n}|+|\tilde \phi^{(r)}_{n}\rangle\langle \phi^{(r)}_{n}|\big),
		\label{eq:eta}
\end{eqnarray}
where $\sigma_n= \pm1$ can be arbitrarily assigned for different $n$.
It can be easily verified that the pseudo-Hermiticity holds with the $\hat H$ and $\hat\eta$ in Eqs. \eqref{eq:specH} and \eqref{eq:eta}.
Apparently, there may be many $\hat \eta$'s satisfying Eq. \eqref{eq:pseudo-Hermitian}. 

The pseudo-Hermiticity does not ensure the reality of the spectrum. Actually, it is neither the sufficient nor the necessary condition for the reality of the spectrum.
It is well known that any Hermitian operator has a real spectrum. Thus, the non-Hermitian operator $\hat H$ with the real spectrum must be related to a Hermitian operator $\hat H_h$ with the same spectrum by a similarity transformation $\hat O$:
\begin{eqnarray}
	\hat H_h&=&\hat O\hat H\hat O^{-1}.
	\label{eq:Hermitian_counterpart}
\end{eqnarray}
The $\hat O$ can be easily constructed using the eigenvectors $|\phi_n\rangle$ and $|\phi_n^{(r/l)}\rangle$ of $\hat H_h$ and $\hat H$, respectively, yielding \cite{Mostafazadeh_2002b_JMP}
\begin{eqnarray}
	\hat O=\sum_n |\phi_n\rangle \langle \phi_n^{(l)}|, ~~\hat O^{-1}=\sum_n |\phi_n^{(r)}\rangle \langle \phi_n|.
\end{eqnarray}
 From Eq. \eqref{eq:Hermitian_counterpart}, the Hermiticity $\hat H_h^\dag=\hat H_h$ requires that
 \begin{eqnarray}
 	(\hat O^{\dag})^{-1}\hat H^\dag\hat O^\dag=\hat O\hat H\hat O^{-1},
 \end{eqnarray}
 and then we have
  \begin{eqnarray}
 	\hat H^\dag=(\hat O^{\dag}\hat O)\hat H(\hat O^\dag\hat O)^{-1},
 \end{eqnarray}
 which is just the $O^\dag O$-pseudo-Hermiticity of $\hat H$.
 This means that the operator with real spectrum must be $O^\dag O$-pseudo-Hermitian.
 The inverse statement can also be easily proved.
Thus, the necessary and sufficient condition for an operator with real spectrum is the $O^\dag O$-pseudo-Hermiticity with $\hat O$ being an invertible linear operator, that is, there must be a special $\eta$-pseudo-Hermiticity of $\hat H$ with $\hat\eta=\hat O^\dag\hat O$ \cite{Ashida_Ueda_AdvPhys_2020}.

As a comparison, $\hat H$ is said to be {\it quasi-Hermitian} \cite{Ashida_Ueda_AdvPhys_2020} if there exists one positive-definite Hermitian (not necessarily invertible) operator $\hat {\xi}$, satisfying
\begin{eqnarray}
	\hat H^\dag \hat{\xi}&=&\hat{\xi}\hat H,
	\label{eq:quasi-Hermitian}
\end{eqnarray}
where $\hat\xi$'s can be expressed using the spectral decomposition as 
\begin{equation}
	\hat\xi =\sum_{n}{\xi_n|\xi _n\rangle \langle \xi _n|}
\end{equation}
with $\xi_n>0$ being the eigenvalue of the orthonormal eigenvector $|\xi_n\rangle$.
The quasi-Hermiticity is a sufficient yet necessary condition for the reality of the spectrum \cite{Ashida_Ueda_AdvPhys_2020}. 
For the invertible case of $\hat\xi$, it means that there exists a Hermitian operator $\hat H_h=\hat S\hat H\hat S^{-1}$ that has the identical spectrum,
where the similarity operator $\hat S$ can be constructed as
\begin{equation}
	\hat S\equiv\sum_{n}{\sqrt{\xi_n}|\xi_n\rangle \langle \xi_n|}.
\end{equation}
The Hermiticity of $\hat H_h$ can be easily proved by noting that
\begin{eqnarray}
	\hat H_h^\dag&=&\hat S^{-1}\hat H^\dag\hat S
	=\hat S^{-1}(\hat H^\dag \hat\xi)\hat\xi^{-1}\hat S
	=\hat S^{-1}( \hat\xi \hat H)\hat S^{-1} \notag\\
	&=&\hat S\hat H\hat S^{-1}
	=\hat H_h,
\end{eqnarray}
where we use the quasi-Hermiticity of $\hat H$ in Eq. \eqref{eq:quasi-Hermitian}.

\section{Symmetries of $\hat H_\text{NH}$}
\label{appendix:eta operation}

In this section, we give the explicit forms of symmetries for the non-Hermitian Hamiltonian $\hat H_\text{NH}$ in the main text when $\chi_+=\chi_-\equiv \chi$:
\begin{equation}
	\hat H_{\rm NH}\equiv\hat{\psi}^\dag H \hat{\psi} =\sum_{k\in \text{BZ}}\hat{\psi}_k^\dag H_k \hat{\psi}_k,
	\label{aeq:Hk}
\end{equation}
where the bases in respective real and $k$ spaces read as
\begin{equation}
	\hat{\psi}^\dag\equiv(\{\hat a^\dag_{j\uparrow}~\hat a^\dag_{j\downarrow}~\hat b^\dag_{j\uparrow}~\hat b^\dag_{j\downarrow}\}), ~
	\hat{\psi}\equiv(\{\hat a_{j\uparrow}~\hat a_{j\downarrow}~\hat b_{j\uparrow}~\hat b_{j\downarrow}\})^T,
\end{equation}
and 
\begin{equation}
\hat{\psi}_k^\dag\equiv(\hat a^\dag_{k\uparrow}~\hat a^\dag_{k\downarrow}~\hat b^\dag_{k\uparrow}~\hat b^\dag_{k\downarrow}),~
	~~ \hat{\psi}_k\equiv(\hat a_{k\uparrow}~\hat a_{k\downarrow}~\hat b_{k\uparrow}~\hat b_{k\downarrow})^T,
\end{equation}
related by the discrete Fourier transform \eqref{eq:dFT}, 
and $H$ and $H_k$ are the corresponding coefficient matrices.

\paragraph*{Time-reversal symmetry.} 
Due to the reality of all parameters in $\hat H_\text{NH}$, the ladder respects the time-reversal symmetry:
\begin{equation}
	\hat T \hat H_{\rm NH} \hat T^{-1}=\hat H_{\rm NH},
	\label{eq:time_reversal}
\end{equation}
where the time-reversal operator $\hat T$ acts just as the complex conjugation on numbers, preserving the operators in real space,
\begin{eqnarray}
	\hat T: \hat a_{j\sigma}^{(\dag)}\rightarrow \hat a_{j\sigma}^{(\dag)}, ~~
	\hat b_{j\sigma}^{(\dag)}\rightarrow \hat b_{j\sigma}^{(\dag)},~~
	i\rightarrow -i,
\end{eqnarray}
and correspondingly in $k$ space, it reverses the sign of $k$ of the operators:
\begin{eqnarray}
	\hat T: \hat a_{k\sigma}^{(\dag)}\rightarrow \hat a_{-k,\sigma}^{(\dag)},~~
	\hat b_{k\sigma}^{(\dag)}\rightarrow \hat b_{-k,\sigma}^{(\dag)}.
\end{eqnarray}
Thus, 
\begin{eqnarray}
	\hat T \hat{\psi}_k^{(\dag)} \hat T^{-1} =\hat{\psi}_{-k}^{(\dag)},
\end{eqnarray}
and the time-reversal symmetry Eq. \eqref{eq:time_reversal} requires that
\begin{equation}
	H_k^*=H_{-k},
	\label{eq:time_reversal_k}
\end{equation}
which ensures that the eigenenergies in $k$ space have the relation $\epsilon_{-k,m}=\epsilon_{kn}^*$, where $\epsilon_{kn}$ means the $n$-th eigenenergy of $H_k$.

\paragraph*{Chiral symmetry.} 
Due to the bipartition of the ladder, it has the chiral symmetry:
\begin{equation}
	\hat C \hat H_{\rm NH} \hat C^{-1}=-\hat H_{\rm NH}
	\label{eq:chiral}
\end{equation}
with the chiral operation
\begin{eqnarray}
	\hat C: \hat a_{j\uparrow}^{(\dag)}\rightarrow \hat a_{j\uparrow}^{(\dag)},\,\hat a_{j\downarrow}^{(\dag)}\rightarrow -\hat a_{j\downarrow}^{(\dag)},
\,\hat b_{j\uparrow}^{(\dag)}\rightarrow -\hat b_{j\uparrow}^{(\dag)}, \,\hat b_{j\downarrow}^{(\dag)}\rightarrow \hat b_{j\downarrow}^{(\dag)}\notag\\
\end{eqnarray}
in real space,
and correspondingly 
\begin{eqnarray}
	\hat C: \hat a_{k\uparrow}^{(\dag)}\rightarrow \hat a_{k\uparrow}^{(\dag)},\,\hat a_{k\downarrow}^{(\dag)}\rightarrow -\hat a_{k\downarrow}^{(\dag)},
\,\hat b_{k\uparrow}^{(\dag)}\rightarrow -\hat b_{k\uparrow}^{(\dag)}, \,\hat b_{k\downarrow}^{(\dag)}\rightarrow \hat b_{k\downarrow}^{(\dag)}\notag\\
\end{eqnarray}
in $k$ space.
Thus, 
\begin{eqnarray}
	\hat C \hat{\psi}_k^\dag \hat C^{-1} =\hat{\psi}_k^\dag C,~~
	\hat C \hat{\psi}_k\hat C^{-1} =C^\dag\hat{\psi}_k,
	\label{eq:sublattice_representation}
\end{eqnarray}
with 
\begin{equation}
	C=
	\left(\begin{array}{cccc}
		1 & 0 & 0  & 0\\
		0 & -1 & 0 & 1\\
		0  & 0 & -1 & 0\\
		0 & 1 & 0 & 1
	\end{array}\right)
	=\tau _z\otimes \sigma _z=C^\dag=C^{-1},
\end{equation}
which indicates that $\hat C^\dag=\hat C^{-1}=\hat C$. 
Hereafter, $\tau_{x,y,z}$ and $\sigma_{x,y,z}$ are both Pauli matrices acting on the sublattice ($a,b$) and the leg ($\uparrow,\downarrow$) spaces, respectively; 
$\tau_0$ and $\sigma_0$ are the respective identity matrices. 
The chiral symmetry Eq. \eqref{eq:chiral} requires that
\begin{equation}
	C H_k C^\dag=-H_k,
	\label{eq:sublattice_k}
\end{equation}
which ensures that the eigenenergies in $k$ space have the relation $\epsilon_{km}=-\epsilon_{kn}$.

\paragraph*{Inversion symmetry.}
The ladder also has the inversion symmetry:
\begin{equation}
	\hat I \hat H_{\rm NH} \hat I^{-1}=\hat H_{\rm NH}
	\label{eq:inversion}
\end{equation}
with the inversion operation
\begin{eqnarray}
	\hat I: \hat a_{j\sigma}^{(\dag)}\rightarrow \hat a_{-j,\bar\sigma}^{(\dag)}, ~~\hat b_{j\sigma}^{(\dag)}\rightarrow \hat b_{-(j+1),\bar\sigma}^{(\dag)},
\end{eqnarray}
in real space, where $\bar\sigma$ means the opposite of leg $\sigma$,
and correspondingly 
\begin{equation}
	\hat I: \hat a_{k\sigma}^{(\dag)}\rightarrow \hat a_{-k,\bar\sigma}^{(\dag)}, ~\hat b_{k\sigma}^{\dag}\rightarrow e^{-ik}\hat b_{-k,\bar\sigma}^{\dag},~\hat b_{k\sigma}\rightarrow e^{ik}\hat b_{-k,\bar\sigma},
\end{equation}
in $k$ space.
Thus, 
\begin{eqnarray}
	\hat I \hat{\psi}_k^\dag \hat I^{-1} =\hat{\psi}_{-k}^\dag I_k,~~
	\hat I \hat{\psi}_k\hat I^{-1} = I_k^\dag\hat{\psi}_{-k},
\end{eqnarray}
with 
\begin{eqnarray}
	I_k&=&
	\left(\begin{array}{cccc}
		0 & 1 & 0 & 0\\
		1 & 0 & 0 & 0\\
		0 & 0 & 0 & e^{-ik}\\
		0 & 0 & e^{-ik} & 0
	\end{array}\right)
	=	\left(\begin{array}{cc}
		1 & 0\\
		0 & e^{-ik}
	\end{array}\right)
	\otimes \sigma _x, \notag\\
	I_k^\dag&=&I_k^{-1}\neq I_k,
\end{eqnarray}
which indicates that $\hat I^\dag=\hat I^{-1}=\hat I$.
The inversion symmetry Eq. \eqref{eq:inversion} requires that
\begin{equation}
	I_k H_k I_k^\dag=H_{-k},
	\label{eq:inversion_k}
\end{equation}
which ensures that the eigenenergies  in $k$ space have the relation $\epsilon_{-k,m}=\epsilon_{kn}$.

\paragraph*{Pseudo-mirror symmetry.}
The ladder also has the pseudo-mirror symmetry:
\begin{equation}
	\hat M \hat H_{\rm NH} \hat M^{-1}=\hat H^\dag_{\rm NH}
	\label{eq:pseudo-mirror}
\end{equation}
with the mirror operation
\begin{eqnarray}
	\hat M: \hat a_{j\sigma}^{(\dag)}\leftrightarrow \hat b_{-j,\sigma}^{(\dag)},
\end{eqnarray}
in real space,
and correspondingly 
\begin{eqnarray}
	\hat M: \hat a_{k\sigma}^{(\dag)}\leftrightarrow \hat b_{-k,\sigma}^{(\dag)},
\end{eqnarray}
in $k$ space.
Thus, 
\begin{eqnarray}
	\hat M \hat{\psi}_k^\dag \hat M^{-1} =\hat{\psi}_{-k}^\dag M,~~
	\hat M \hat{\psi}_k\hat M^{-1} = M\hat{\psi}_{-k},
\end{eqnarray}
with 
\begin{equation}
	M=
	\left(\begin{array}{cccc}
		0 & 0 & 1 & 0\\
		0 & 0 & 0 & 1\\
		1 & 0 & 0 & 0\\
		0 & 1 & 0 & 0
	\end{array}\right)
	=\tau _x\otimes \sigma _0=M^\dag=M^{-1},
\end{equation}
which indicates that $\hat M^{\dag}=\hat M^{-1}=\hat M$.
The pseudo-mirror symmetry Eq. \eqref{eq:pseudo-mirror} requires that
\begin{equation}
	M H_k M=H_{-k}^\dag,
	\label{eq:pseudo-mirror_k}
\end{equation}
which ensures that the eigenenergies in $k$ space have the relation $\epsilon_{-k,m}=\epsilon_{kn}^*$.
This symmetry can be regarded as a non-Hermitian generalization of the traditional mirror symmetry in Hermitian systems.

\paragraph*{Pseudo-glide symmetry.}
The ladder also has the pseudo-glide symmetry:
\begin{equation}
	\hat G \hat H_{\rm NH} \hat G^{-1}=\hat H_{\rm NH}^\dag
	\label{eq:pseudo-glide}
\end{equation}
with the pseudo-glide operation:
\begin{eqnarray}
	\hat G: \hat a_{j\sigma}^{(\dag)}\rightarrow \hat b_{j\bar\sigma}^{(\dag)}, ~~\hat b_{j\sigma}^{(\dag)}\rightarrow \hat a_{j+1,\bar\sigma}^{(\dag)},
\end{eqnarray}
in real space, which also satisfies the relation $\hat G=\hat M\hat I$, 
and correspondingly 
\begin{equation}
	\hat G: \hat a_{k\sigma}^{(\dag)}\rightarrow \hat b_{k\bar\sigma}^{(\dag)}, ~\hat b_{k\sigma}^{\dag}\rightarrow e^{-ik}\hat a_{k\bar\sigma}^{\dag}, ~\hat b_{k\sigma}\rightarrow e^{ik}\hat a_{k\bar\sigma},
\end{equation}
in $k$ space.
Thus, 
\begin{eqnarray}
	\hat G \hat{\psi}_k^\dag \hat G^{-1} =\hat{\psi}_{k}^\dag G_k,~~
	\hat G \hat{\psi}_k\hat G^{-1} = G_k^\dag\hat{\psi}_{k},
\end{eqnarray}
with 
\begin{eqnarray}
	G_k&=&
	\left(\begin{array}{cccc}
		0 & 0 & 0 & e^{-ik}\\
		0 & 0 & e^{-ik} & 0\\
		0 & 1 & 0 & 0\\
		1 & 0 & 0 & 0
	\end{array}\right)
	=	\left(\begin{array}{cc}
		0 & e^{-ik}\\
		1 & 0
	\end{array}\right)
	\otimes \sigma _x
	=MI_k, \notag\\
	G_k^\dag&=&G_k^{-1}\neq G_k, ~G_k^2= e^{-ik} I ~(I: \text{identity matrix}),\notag\\
\end{eqnarray}
which indicates that $\hat G^\dag=\hat G^{-1}\ne\hat G$.
The pseudo-glide symmetry Eq. \eqref{eq:pseudo-glide} requires that
\begin{equation}
	G_k H_k G_k^\dag=H^\dag_{k},
	\label{eq:pseudo-glide_k}
\end{equation}
which ensures that the eigenenergies in $k$ space have the relation $\epsilon_{km}=\epsilon_{kn}^*$.
This symmetry can be regarded as a non-Hermitian generalization of the traditional glide symmetry in Hermitian systems.

\paragraph*{Pseudo-Hermitian symmetry.}

The pseudo-glide relation \eqref{eq:pseudo-glide_k} is close to the form of the pseudo-Hermiticity \eqref{eq:pseudo-Hermitian}, with only the violation that $G_k$ is not Hermitian. 
To make up this incompleteness, we can define 
\begin{eqnarray}
	\eta_k &\equiv& e^{ik/2}G_k 
	=\left(\begin{array}{cccc}
		0 & 0 & 0 & \mathrm{e}^{-\mathrm{i}k/2} \\
		0 & 0 & \mathrm{e}^{-\mathrm{i}k/2} & 0\\
		0 & \mathrm{e}^{\mathrm{i}k/2}  & 0 & 0\\
		\mathrm{e}^{\mathrm{i}k/2} & 0 & 0 & 0
	\end{array}\right)\notag\\
	&=&\left(\begin{array}{cc}
		0 & e^{-ik/2}\\
		e^{ik/2} & 0
	\end{array}\right)\otimes\sigma_x
	=\eta_k^\dag=\eta_k^{-1}
\end{eqnarray}
such that 
\begin{equation}
	\eta_k H_k \eta_k^{-1}=H_k^\dag
\end{equation}
with a Hermitian matrix $\eta_k$, that is, $H_k$ is $\eta_k$-pseudo-Hermitian.
Assuming that $\hat \eta$ is the pseudo-Hermitian operator, to make $\hat H_\text{NH}$ $\eta$-pseudo-Hermitian, i.e., 
\begin{equation}
	\hat \eta\hat H_\text{NH}\hat \eta^{-1}=\hat H_\text{NH}^\dag,
\end{equation} 
we have
\begin{equation}
	\hat \eta \hat \psi_k^\dag \hat \eta^{-1}=\hat\psi_k^\dag\eta_k,~~ \hat \eta \hat \psi_k \hat \eta^{-1}=\eta_k^\dag\hat\psi_k,
\end{equation}
and thus,
\begin{eqnarray}
	\hat \eta:~&& \hat a_{k\sigma}^{\dag}\rightarrow e^{ik/2}\hat b_{k\bar\sigma}^{\dag}, ~~~\hat b_{k\sigma}^{\dag}\rightarrow e^{-ik/2}\hat a_{k\bar\sigma}^{\dag},\notag\\
	&&\hat a_{k\sigma}\rightarrow e^{-ik/2}\hat b_{k\bar\sigma}, ~~\hat b_{k\sigma}\rightarrow e^{ik/2}\hat a_{k\bar\sigma},
\end{eqnarray}
in $k$ space, which indicates that $\hat \eta^\dag=\hat \eta^{-1}=\hat \eta$.
Correspondingly, using the discrete Fourier transform \eqref{eq:dFT}, we obtain the operations in real space:
\begin{eqnarray}
	\hat \eta:~&& \hat a_{j\sigma}^{\dag}\rightarrow\sum_{j'}\hat b_{j'\bar\sigma}^{\dag}\eta_{j'j},~~
	\hat b_{j\sigma}^{\dag}\rightarrow\sum_{j'}\hat a_{j'\bar\sigma}^{\dag}\eta_{j'j}^*,\notag\\
	&&\hat a_{j\sigma}\rightarrow\sum_{j'}\hat b_{j'\bar\sigma}\eta_{j'j}^*,~~
	\hat b_{j\sigma}\rightarrow\sum_{j'}\hat a_{j'\bar\sigma}\eta_{j'j},
\end{eqnarray}
where 
\begin{eqnarray}
	\eta_{j'j}&=&\frac{1}{L}\sum_k e^{-ik(j-j'-1/2)} \notag\\
	&=&\frac{2L^{-1}}{1-\exp[-i2\pi L^{-1}(j-j'-1/2)]}.
\end{eqnarray}
This expression clearly demonstrates that the pseudo-Hermiticity operation is highly complex, transforming a single site of a leg to a superposition of all sites of another leg, not like the common operations such as chiral, mirror, inversion, and glide operations, which only involves single-site transformations.

Compared with the Hamiltonian $\hat H$ of the Hermitian SSH ladder, the dissipation-induced nonreciprocal hopping in $\hat H_\text{NH}$ relaxes some traditional symmetries, such as the mirror and the glide symmetries, to the corresponding pseudo ones, and also generalizes the time-reversal symmetry to the field of complex energies.

When $\chi_+\ne \chi_-$, one can follow the similar procedure to verify that the time-reversal, the chiral, and the inversion symmetries are preserved, but the pseudo-mirror, the pseudo-glide, and thus the $\eta_k$-pseudo-Hermitian symmetries are broken. 
Even in this case, the preserved symmetries are sufficient to ensure the $D_{2h}$ symmetry of the band structure in the composite 3D space spanned by the three twofold rotating axies: the $k$ axis, the real and the imaginary axes of the complex energy plane.
This symmetry implies that the ladder possesses other hidden pseudo-Hermitian symmetry instead of the $\eta_k$-pseudo-Hermitian symmetries in Eq. \eqref{eq:pseudo-Hermitian_k} according to the statement in Appendix \ref{asec:recap}, although the explicit form of the pseudo-Hermitian operator cannot be easily written.

\section{A gauge-invariant winding number for non-Hermitian multi-band systems under the chiral symmetry}
\label{asec:winding_number}

Due to the chiral symmetry $\{H_k, C\}=0$, the right or left eigenvectors of the Hamiltonian matrix $H_k$ of dimension $2N$ appear in pairs for each $k$, $\{Cu^{(r/l)}_{kn}, u^{(r/l)}_{kn}\}$, with the corresponding sign-inverted eigenenergies $\{\epsilon_{kn},-\epsilon_{kn}\}$. 
Thus, we can categorize these eigenvectors into two subspaces $\mathbb{G}_\pm $ of dimension $N$, and define a $2N\times N$ matrix $U_-^{(r/l)}\equiv(u_{k1}^{(r/l)},\dots,u_{kN}^{(r/l)})$ to group the eigenvectors in $\mathbb{G}_-$ and the matrix $U_+^{(r/l)}\equiv CU_-^{(r/l)}$ in $\mathbb{G}_+$, where the subscripts ``$\pm$'' are simply labels, as the eigenenergies are generally complex for non-Hermitian Hamiltonians. One can choose eigenvectors of the subspaces for demand, without necessarily ordering them by the real or imaginary parts of the eigenenergies. 
Hereafter, we assume that the defectiveness (i.e., non-diagonalization) of the non-Hermitian Hamiltonian $H_k$ occurs only at finite discrete sets of system parameters such that the biorthogonal theory can be analytically used at the neighborhood of these points, preventing using the theory of generalized eigenvectors.

Given the biorthonormal relations between right and left eigenvectors, i.e., $U_\pm^{(l)\dag}U_\pm^{(r)}=I_{N}$ and $U_\pm^{(l)\dag}U_\mp^{(r)}=0$, where $I_{N}$ is the identity matrix of dimension $N$, we can define a $2N\times 2N$ matrix $U_k\equiv (U_+^{(r)},U_-^{(r)})$ to diagonalize $H_k$ as
\begin{eqnarray}
	U_k^{-1}H_kU_k
	=	\left(
	\begin{array}{cc}
		E_k & 0  \\
		0 & -E_k 
	\end{array}\right),
\end{eqnarray}
where $U_k^{-1}=(U_+^{(l)},U_-^{(l)})^\dag$ and $E_k\equiv\text{diag}(\epsilon_{k1},\dots,\epsilon_{kN})$.
Thus, we can rewrite $H_k$ using the eigenvectors and the eigenenergies as
\begin{eqnarray}
	H_k=U_+^{(r)}E_kU_+^{(l)\dag}-U_-^{(r)}E_kU_-^{(l)\dag}.
\end{eqnarray}

Due to the chiral symmetry $\{H_k,C\}=0$, $H_k$ can be brought into a block off-diagonal form:
\begin{equation}
	H_k^{(b)}\equiv U_s^{-1}H_kU_s
	\equiv\left(\begin{array}{cc}
		0 & h_k^{(+)} \\
		h_k^{(-)} & 0 
	\end{array}\right)
\end{equation}
with two $N\times N$ matrices $h_k^{(+)}$ and $h_k^{(-)}$, under the basis represented by the unitary matrix $U_s$ ($U_s$-basis) that diagonalizes $C$, i.e., 
\begin{equation}
	U_s^{\dag}CU_s=
	\left(\begin{array}{cc}
		I_{N} & 0  \\
		0 & -I_{N} 
	\end{array}\right).
\end{equation}
This can be proven as follows: 
\begin{equation}
	0 = U_s^{\dag}\{H_k,C\}U_s=
	\left\{H^{(b)}_k,
	\left(\begin{array}{cc}
		I_{N} & 0  \\
		0 & -I_{N} 
	\end{array}\right)\right\},
\end{equation}
which leads to
\begin{equation}
	H^{(b)}_k=
	\left(\begin{array}{cc}
		-I_{N} & 0  \\
		0 & I_{N} 
	\end{array}\right)H_k^{(b)} 
	\left(\begin{array}{cc}
		I_{N} & 0  \\
		0 & -I_{N} 
	\end{array}\right).
\end{equation}
This relation clearly shows that $H^{(b)}_k$ must be off-diagonal. 

To define the topology, one can construct a $Q$-matrix \cite{Ryu_Ludwig_2010_NJP} 
that has identical eigenvectors but with eigenvalues $\pm E_k$ of the eigenvectors in $\mathbb{G}_\pm$ subspaces collapsed to $\pm 1$, respectively: 
\begin{eqnarray}
	Q_k&\equiv&U_+^{(r)}U_+^{(l)\dag}-U_-^{(r)}U_-^{(l)\dag}.
	\label{eq:Qmatrix}
\end{eqnarray}
It can be easily verified that $Q_k$ shares the same chiral symmetry as $H_k$, i.e., $\{Q_k,C\}=0$. 
Likewise, $Q_k$ can also be cast in a block off-diagonal form under the same $U_s$-basis of $H_k^{(b)}$:
\begin{equation}
	Q_k^{(b)}\equiv U_s^{\dag}Q_kU_s
	\equiv\left(\begin{array}{cc}
		0 & q_k  \\
		q^{-1}_k & 0 
	\end{array}\right),
	\label{eq:Qmatrix_block}
\end{equation}
where $q_k$ is an $N\times N$ invertible matrix (i.e., $\det q_k\neq 0$). That the two block matrices are invertible to each other is derived from $[Q_k^{(b)}]^2=Q_k^2=I_{2N}$.
Then, we can define the winding number:
\begin{equation}
	w \equiv \frac{1}{2\pi i}\int_{k \in \text{BZ}} \text{tr}\,q_k^{-1}\text{d} q_k
	=\frac{1}{2\pi i}\int_{k \in \text{BZ}} \text{d} \ln\det q_k.
	\label{eq:winding}
\end{equation}
Here, the second identity can be obtained using the following relation:
\begin{eqnarray}
	\text{tr}\,q_k^{-1}\text{d}q_k
	&=&\text{tr}\,U\Lambda^{-1}U^{-1} \text{d}\,U\Lambda U^{-1} 
	= \text{tr}\,\Lambda^{-1}\text{d}\Lambda \notag\\
	&=& \text{d\,tr}\ln\Lambda
	= \text{d}\ln\det \Lambda 
	= \text{d}\ln\det U\Lambda U^{-1} \notag\\
	&=& \text{d}\ln\det q_k,
	\label{eq:trick}
\end{eqnarray}
where we use the eigenvalue decomposition $U^{-1}q_kU=\Lambda\equiv \text{diag}(\{\lambda_s\})$ with $\lambda_s\,(s=1,\dots,N)$ being the eigenvalues, and $\ln \Lambda\equiv \text{diag}(\{\ln\lambda_s\})$.
This winding number $w$ defines a map from $k\in S^1$ onto an $N\times N$ invertible matrix $q_k\in GL(N,\mathbb{C})$ to reflect the topology of the bipartition of the Hilbert space into two subspaces $\mathbb{G}_\pm$ in BZ. 
This map is a fundamental homotopy group $\pi_1(GL(N,\mathbb{C}))\in \mathbb{Z}$.
Specifically, when $H_k$ and thus $Q_k$ and $Q_k^{(b)}$ are Hermitian, the invertible matrix $q_k$ becomes unitary (i.e., $q_k^\dag=q_k^{-1}$), reducing the fundamental homotopy group from $\pi_1(GL(N,\mathbb{C}))$ to $\pi_1(U(N))\in \mathbb{Z}$.

It is important to note that given the $U_s$-basis, which changes $q_k$ but preserves $w$ due to the independence of $U_s$ on $k$, the block matrix $q_k$ and thus the winding number $w$ are invariant with respect to any local (i.e., $k$-dependent) invertible linear transformation $R_k$ of the eigenvectors in each subspace,
i.e., $\tilde U_\pm^{(r)}=U_\pm^{(r)}R_k$ and $\tilde U_\pm^{(l)\dag}=R_k^{-1}U_\pm^{(l)\dag}$.
This gauge invariance reflects the robustness of the winding number $w$, provided that the two sets of bands do not touch for any $k$ in BZ.
Apparently, this gauge invariance does not hold for $h_k^{(+)}$ or $h_k^{(-)}$, and thus, the winding number cannot be properly defined using them.

To explicitly express $q_k$, we rewrite the eigenvectors under the $U_s$-basis as $U_s^{\dag}U_-^{(r/l)}\equiv (U_1^{(r/l)};U_2^{(r/l)})$, where the two $N\times N$ matrices $U_{1,2}^{(r/l)}$ are arranged in a vertical order, denoted by ``;" in the parentheses.
According to the biorthonormal and completeness relations:
\begin{eqnarray}
	I_N&=&[U_-^{(l)\dag}U_s][U_s^\dag U_-^{(r)}]
	=U_1^{(l)\dag}U_1^{(r)}+U_2^{(l)\dag}U_2^{(r)}, \notag\\
	0&=&[U_+^{(l)\dag}U_s][U_s^\dag U_-^{(r)}]
	=U_1^{(l)\dag}U_1^{(r)}-U_2^{(l)\dag}U_2^{(r)}, \notag\\
	I_{2N}&=&U_s^\dag [U_+^{(r)} U_+^{(l)\dag}+U_-^{(r)} U_-^{(l)\dag}]U_s \notag\\
	&=& \left(\begin{array}{cc}
		2U_1^{(r)}U_1^{(l)\dag} & 0  \\
		0 & 2U_2^{(r)}U_2^{(l)\dag} 
	\end{array}\right),
\end{eqnarray}
one can immediately obtain
\begin{eqnarray}
	U_{1,2}^{(l)\dag}U_{1,2}^{(r)}
	=U_{1,2}^{(r)}U_{1,2}^{(l)\dag}
	=\frac{I_N}{2}.
\end{eqnarray}
From Eqs. \eqref{eq:Qmatrix} and \eqref{eq:Qmatrix_block}, one has
\begin{eqnarray}
	Q_k^{(b)}&=&U_s^\dag [CU_-^{(r)}U_-^{(l)\dag}C-U_-^{(r)}U_-^{(l)\dag}]U_s \notag\\
&=&
\left(\begin{array}{cc}
	I_{N} & 0  \\
	0 & -I_{N} 
\end{array}\right)
U_s^\dag U_-^{(r)}U_-^{(l)\dag}U_s
\left(\begin{array}{cc}
	I_{N} & 0  \\
	0 & -I_{N} 
\end{array}\right) \notag\\
&& -U_s^\dag U_-^{(r)}U_-^{(l)\dag}U_s \notag\\
&=&
\left(\begin{array}{cc}
	0 & -2U_1^{(r)}U_2^{(l)\dag}  \\
	-2U_2^{(r)}U_1^{(l)\dag} & 0 
\end{array}\right).
\label{eq:qk_derive}
\end{eqnarray}
Thus, 
\begin{equation}
	q_k=-2U_1^{(r)}U_2^{(l)\dag},~~q_k^{-1}=-2U_2^{(r)}U_1^{(l)\dag},
	\label{eq:q_expression}
\end{equation}
which are apparently gauge invariant.
These expressions offer a way of numerically calculating the winding number \eqref{eq:winding} with eigenvectors. 

Using these expressions of $q_k$ and $q_k^{-1}$ as well as the above relations, the winding number can be further calculated as
\begin{eqnarray}
	w &=& \frac{1}{2\pi i}\int_{k \in \text{BZ}} \text{tr}\,q_k^{-1}\text{d} q_k \notag\\
	&=& \frac{1}{\pi i}\int_{k \in \text{BZ}} \text{tr}\,[U_1^{(l)\dag}\text{d} U_1^{(r)}- U_2^{(l)\dag}\text{d}U_2^{(r)}] \notag\\
	&=& \frac{1}{\pi i}\int_{k \in \text{BZ}} \text{tr}\,U_+^{(l)\dag}\text{d} U_-^{(r)} \notag\\
	&=&\frac{1}{\pi i}\int_{k \in \text{BZ}} \text{tr}\,U_-^{(l)\dag}C\text{d} U_-^{(r)},
\end{eqnarray}
which can be regarded as an alternative definition of the gauge-independent winding number using eigenvectors in two subspaces.
The $C$ matrix is the key to immune to the gauge change, which leads to the essential difference to the winding number defined using eigenvectors in only one subspace, say
\begin{eqnarray}
	w_- &\equiv&\frac{1}{\pi i} \int_{k \in \text{BZ}} \text{tr}\, U_-^{(l)\dag}\text{d} U_-^{(r)}\notag\\
	&=&  \frac{1}{\pi} \int_{k \in \text{BZ}} \text{tr}\, A_k^{(-)} 
	=\frac{\varphi_{\text Zak}}{\pi}.
	\label{aeq:Zak}
\end{eqnarray}
Here, $\varphi_\text{Zak}\equiv \int_{k \in \text{BZ}} \text{tr}\, A_k^{(-)} $ is the Zak phase defined in BZ for the subspace $\mathbb{G}_-$ using the non-Abelian Berry connection $A_k^{(-)} \equiv-iU_-^{(l)\dag}\text{d} U_-^{(r)}$.
This definition is not gauge invariant, since
\begin{eqnarray}
	&&\tilde w_- = \frac{1}{\pi i} \int_{k \in \text{BZ}} \text{tr}\, \tilde U_-^{(l)\dag}\text{d} \tilde U_-^{(r)} \notag\\
	&=& w_- +\frac{1}{\pi i} \int_{k \in \text{BZ}} \text{tr}\, R_k^{-1}\text{d} R_k
	=w_- +\frac{1}{\pi i} \int_{k \in \text{BZ}} \text{tr}\, \text{d}\ln \Lambda \notag\\
	&=&w_- +\frac{1}{\pi i} \sum_{n=1}^{N}\int_{k \in \text{BZ}} (\text{d}\ln r_n +i\text{d}\varphi_n) \notag\\
	&=&w_- +\frac{1}{\pi} \sum_{n=1}^{N}\int_{k \in \text{BZ}} \text{d}\varphi_n
	=  w_- +2\times \text{Integer},
	\label{eq:gauge_variance}
\end{eqnarray}
where we use the same trick as in Eq. ({\ref{eq:trick}}), $U^{-1}R_kU=\Lambda\equiv \text{diag} (\{r_n e^{i\varphi_n}\})~(n=1,\dots,N)$ with $r_n\ge 0$ and $\varphi_n\in \mathbb{R} $, and notice that $r_n$ is periodic in $k$ but $\varphi_n$ is periodic modulo $2\pi$, since $R_k$ is a periodic function of $k$.
Equation \eqref{eq:gauge_variance} demonstrates that this definition of winding number is only valid modulo $2$, i.e., $w_- \in \mathbb{Z}_2$, much coarser than $w\in \mathbb{Z}$.
It is interesting to note that the two kinds of winding numbers come from different parts of the matrix of the non-Hermitian non-Abelian Berry connection for the whole bands:
\begin{equation}
	A_k\equiv -iU_k^{-1}\mathrm{d}U_k
	=\left(\begin{array}{cc}
		-iU_+^{(l)\dag}\mathrm{d}U_+^{(r)} & -iU_+^{(l)\dag}\mathrm{d}U_-^{(r)}  \\
		-iU_-^{(l)\dag}\mathrm{d}U_+^{(r)} & -iU_-^{(l)\dag}\mathrm{d}U_-^{(r)} 
	\end{array}\right).
	\label{eq:non-Abelian_BC}
\end{equation}

From the point of view of $H_k$, one can also construct $H_k^{(b)}$ as in Eq. \eqref{eq:qk_derive}:
\begin{eqnarray}
	H_k^{(b)}&=&U_s^\dag [CU_-^{(r)}E_kU_-^{(l)\dag}C-U_-^{(r)}E_kU_-^{(l)\dag}]U_s \notag\\
		&=&
	\left(\begin{array}{cc}
		0 & -2U_1^{(r)}E_kU_2^{(l)\dag}  \\
		-2U_2^{(r)}E_kU_1^{(l)\dag} & 0 
	\end{array}\right).
\end{eqnarray}
Thus, 
\begin{equation}
	h_k^{(+)}=-2U_1^{(r)}E_kU_2^{(l)\dag},~~h_k^{(-)}=-2U_2^{(r)}E_kU_1^{(l)\dag},
	\label{eq:h_expression}
\end{equation}
which apparently are both gauge dependent.
Combined with Eq. \eqref{eq:q_expression},  we have the relations between $q_k$ and $h_k^{(\pm)}$:
\begin{equation}
	q_k=h_k^{(+)}[U_2^{(l)\dag}]^{-1}E_k^{-1}U_2^{(l)\dag}
	=[U_1^{(l)\dag}]^{-1}E_kU_1^{(l)\dag}h_k^{\prime -1},
\end{equation}
and thus, the winding number \eqref{eq:winding} can be reexpressed as
\begin{eqnarray}
	w &=&\frac{1}{2\pi i}\int_{k \in \text{BZ}} \text{d} \ln\det q_k \notag\\
	&=&\frac{1}{2\pi i}\int_{k \in \text{BZ}} \text{d} \ln\det h_k^{(+)}-\frac{1}{2\pi i}\int_{k \in \text{BZ}} \text{d} \ln\det E_k \notag\\
	&=&-\frac{1}{2\pi i}\int_{k \in \text{BZ}} \text{d} \ln\det h_k^{(-)}+\frac{1}{2\pi i}\int_{k \in \text{BZ}} \text{d} \ln\det E_k. \notag\\
	\label{eq:winding_h}
\end{eqnarray}
These expressions demonstrate that if one tries to calculate the gauge-invariant winding number using either $h_k^{(+)}$ or $h_k^{(-)}$, the complex eigenenergies $E_k$ of the non-Hermitian systems must be involved simultaneously, which in Hermitian systems does not contribute due to the reality of energies and reduces to the traditional expression.
For the energy spectrum that respects the mirror symmetry for each $k$, there always exist counter-rotating traces of eigenenergies along with $k$, and thus the energy term in Eq. \eqref{eq:winding_h} must vanish;
for the energy spectrum separated by a line gap, the energy term also vanishes because the origin of the complex energy plane is not enclosed; 
for the energy spectrum separated by a point gap enclosing the origin but not respecting the mirror symmetry, the energy term must be carefully considered.
Note that although $h_k^{(\pm)}$ is gauge dependent, $\det h_k^{(\pm)}$ is not and thus, the $h_k^{(\pm)}$ part in Eq. \eqref{eq:winding_h} is still gauge invariant, which can be used as another definition of gauge-invariant winding numbers:
 \begin{eqnarray}
 	w_h^{(\pm)} &\equiv& \frac{1}{2\pi i}\int_{k \in \text{BZ}} \text{tr}\,[h_k^{(\pm)}]^{-1}\text{d} h_k^{(\pm)} \notag\\
 	&=&\frac{1}{2\pi i}\int_{k \in \text{BZ}} \text{d} \ln\det h_k^{(\pm)}.
 \end{eqnarray} 
The winding number $w_h^{(\pm)}$ also defines a map from $k\in S^1$ onto an $N\times N$ invertible matrix $h_k^{(\pm)}\in GL(N,\mathbb{C})$, also corresponding to a fundamental homotopy group $\pi_1(GL(N,\mathbb{C}))\in \mathbb{Z}$.
The difference between $w$ and $w_h^{(\pm)}$ is whether considering the contribution from the winding of the complex eigenenergies.

To easily calculate $w$ with only the system parameters $h_k^{(\pm)}$, the winding number can be rewritten in a more delicate way by adding the last two lines of Eq. \eqref{eq:winding_h}, yielding
\begin{eqnarray}
	w &=&\frac{1}{4\pi i}\int_{k \in \text{BZ}}  \left[\text{d}\ln\det h_k^{(+)}-\text{d}\ln\det h_k^{(-)}\right]
	\notag\\
	&=&\frac{1}{2}[w_h^{(+)}-w_h^{(-)}].
\end{eqnarray}
Two useful relations are listed here:
\begin{equation}
	\frac{\det h_k^{(+)}}{\det h_k^{(-)}}=(\det q_k)^2,~\det h_k^{(+)}\det h_k^{(-)}=(\det E_k)^2
\end{equation}
for understanding the properties of this winding number.

\section{Relation between the dynamics and the Wilson line for non-Hermitian systems}
\label{asec:wilson_line}

When we add an external constant force $F$ along the ladder, i.e., the Hamiltonian in Eq. \eqref{eq:Lindblad} becomes $\hat H-F\hat X$, 
where for simplicity the position operator $\hat X$ is defined only with respect to the positions of unit cells, i.e.,
	\begin{eqnarray}
		\hat X=\sum_{j\sigma} j(\hat a_{j\sigma}^{\dag}\hat a_{j\sigma}+\hat b_{j\sigma}^{\dag}\hat b_{j\sigma}).
	\end{eqnarray}
Following the similar derivation from Eq. \eqref{eq:master_Heff} to \eqref{eq:effective Ham}, the short-time dynamics before a quantum jump occurs can be well captured by the time-dependent non-Hermitian Schr\"odinger equation ($\hbar=1$):
	\begin{equation}
		i{\partial_t} |\psi (t)\rangle = (\hat H_\text{NH}-F \hat X)|\psi (t)\rangle,
	\label{eq:Schrodinger}
	\end{equation}
where $\partial_t\equiv \partial/\partial t$, and the loss term $-i\delta\hat N$ in Eq. \eqref{eq:effective Ham}, which only contributes an overall decaying factor to the dynamics, is ignored.
Here, we rewrite the Hamiltonian \eqref{aeq:Hk} in the diagonal form:
\begin{equation}
	\hat H_\text{NH}=\sum_{k\in{\text BZ}}\hat u_k^{(r)\dag}\Lambda_k\hat u^{(l)}_k
	=\sum_{k\in{\text BZ}}\sum_{n=1}^4\epsilon_{kn}\hat u_{kn}^{(r)\dag}\hat u^{(l)}_{kn},
\end{equation}
where $\hat u_k^{(r)\dag}\equiv[\hat u_{k1}^{(r)\dag},\hat u_{k2}^{(r)\dag},\hat u_{k3}^{(r)\dag},\hat u_{k4}^{(r)\dag}]=\hat \psi_k^\dag U_k$, 
$\hat u_k^{(l)}\equiv[\hat u_{k1}^{(l)},\hat u_{k2}^{(l)},\hat u_{k3}^{(l)},\hat u_{k4}^{(l)}]^T=U_k^{-1}\hat \psi_k$, 
and 
$\Lambda_k=\text{diag}(\epsilon_{k1},\epsilon_{k2},\epsilon_{k3},\epsilon_{k4})$.
We set the initial state as a superposition of Bloch states of different bands at $k_0$, i.e., 
$|\psi(0)\rangle = \hat u_k^{(r)\dag} \alpha(0) |0\rangle$,
where the coefficient vector $\alpha(t)=[\alpha_1(t),\alpha_2(t),\alpha_3(t),\alpha_4(t)]^T$ with $\alpha_n(t)$ being the amplitude of Bloch state in the $n$th band at time $t$, and we normalize the coefficients at $t=0$ with $\sum_n|\alpha_n(0)|^2=1$.
With the ansatz that the initial state uniformly evolves to another superposition of Bloch states 
at $k(t) = k_0 + Ft$, i.e., $|\psi(t)\rangle \approx  \hat u_{k(t)}^{(r)\dag} \alpha(t) |0\rangle$ \cite{Li_Schneider_2016_Science}, Eq. \eqref{eq:Schrodinger} reduces to
\begin{equation}
	i{\partial_t} \alpha(t)= [\Lambda_{k(t)}+FA_{k(t)}]\alpha(t),
	\label{eq:dynamics}
\end{equation}
where 
\begin{equation}
A_{k(t)}=-iU_k^{-1}\partial_k U_k\big|_{k=k(t)}
\end{equation}
with $\partial_k\equiv\partial/\partial_k$ is just the non-Hermitian non-Abelian Berry connection Eq. \eqref{eq:non-Abelian_BC} for the whole bands.
The solution to Eq. \eqref{eq:dynamics} is 
\begin{eqnarray}
	\alpha(t)&=& \mathcal{T}\exp\Big\{-i\int_0^t dt\big[\Lambda_{k(t)}+FA_{k(t)}\big]\Big\}\alpha(0) \notag \\
	&=& \mathcal{P}\exp\Big\{-i\int_{k_0}^{k_f} dk\big(\Lambda_{k}/F+A_{k}\big)\Big\}\alpha(0)
	\notag\\
	&\equiv& V_{k_0\rightarrow k_f} \alpha(0),
	\label{eq:dynamics_sol}
\end{eqnarray}
where $\mathcal{T}$ and $\mathcal{P}$ are the time-ordered and the path-ordered operators, respectively.

To numerically solve this equation, we divide the path $k_0\rightarrow k_f$ into $l$ pieces of equal intervals of length $dk=(k_f-k_0)/l$. 
When $l\rightarrow\infty$ (i.e., $dk\rightarrow 0$), using the relation
\begin{eqnarray}
	&&e^{-i (\Lambda_k/F+A_k) d k}\approx e^{-i \Lambda_k dk/F}e^{-iA_k d k}
	\notag\\
	&\approx& e^{-i \Lambda_k dk/F}(I-i A_k d k) 
=e^{-i \Lambda_k dk/F}(I-U_k^{-1}d U_k)
	\notag\\
	&=&e^{-i \Lambda_k dk/F}(I+dU_k^{-1}U_{k} )
    \notag\\
	&\approx& e^{-i \Lambda_k dk/F}[I+(U_{k+d k}^{-1}-U_{k}^{-1})U_k]
	\notag\\
	&=&e^{-i \Lambda_k dk/F}U_{{k}+d {k}}^{-1}U_k,
\end{eqnarray}
we have
\begin{eqnarray}
	V_{k_0\rightarrow k_f}&=&e^{-i \Lambda_{k_f} dk/F}U_{k_f}^{-1} {U}_{k_{l-1}}\dots
	e^{-i \Lambda_{k_i} dk/F}{U}_{{k}_i}^{-1} {U}_{{k}_{i-1}}
	\notag\\ 
	&& \dots e^{-i \Lambda_{k_1} dk/F}{U}_{{k}_1}^{-1} {U}_{{k}_0}.
	\label{eq:V-line}
\end{eqnarray}
where $(k_1,\dots,k_{l-1})$ are the $l-1$ joint points of the $l$ pieces of equal intervals along the path.

When $ F$ is much larger than the total band width $w_4$, the non-Hermitian non-Abelian Berry connection $A_k$ dominates the evolution (i.e., $\Lambda_k/F\rightarrow0$), yielding
\begin{equation}
	\alpha(t)\approx  \mathcal{P}\exp\Big[-i\int_{k_0}^{k_f} dk A_{k}\Big]\alpha(0) 
	\equiv  W_{k_0\rightarrow k_f}\alpha(0),
	\label{aeq:solution}
\end{equation}
where 
\begin{equation}
	W_{k_0\rightarrow k_f}=U_{k_f}^{-1} {U}_{k_{l-1}}\dots {U}_{{k}_i}^{-1} {U}_{{k}_{i-1}} \dots {U}_{{k}_1}^{-1} {U}_{{k}_0}
	=U_{k}^{-1} U_{k_0}.
\end{equation}
is a four-band Wilson line for non-Hermitian systems along the path from $k_0$ to $k_f$ in $k$ space \cite{Xu-Zhu-2022}, reduced from $V_{k_0\rightarrow k_f}$ in Eq. \eqref{eq:V-line}. 
This Wilson line defined using the biorthonormal basis is a non-Hermitian generalization of the Wilson line in Hermitian systems \cite{Li_Schneider_2016_Science,Zhang_Zhou_2017_PRA}. 
Apparently, $W_{k_f\rightarrow k_0}=W_{k_0\rightarrow k_f}^{-1}$.
If $k_f=k_0+2\pi n$ ($n\in$ Integer), the Wilson line becomes a Wilson loop due to the periodicity of the BZ, and thus, $W_{k_0\rightarrow k_0+2\pi n}=I$, indicating the state returns to the initial state, i.e., $|\psi(t)\rangle=|\psi(0)\rangle$.

When the force $F$ is much larger than the maximum energy difference $w_2$ of the lowest two bands and smaller than the gap $\Delta$ to the other two bands, Eq. \eqref{eq:dynamics_sol} can be approximately decoupled into two independent evolutions in each subspace: 
\begin{eqnarray}
	\alpha^{(\pm)}(t)&\approx&  \mathcal{P}\exp\Big[-i\int_{k_0}^{k_f} dk A^{(\pm)}_{k}\Big]\alpha^{(\pm)}(0) \notag
	\\
	&\equiv&  W^{(\pm)}_{k_0\rightarrow k_f}\alpha^{(\pm)}(0),
\end{eqnarray}
where $(\pm)$ means the upper/lower $2\times 2$ diagonal blocks in the labeled matrices and $2\times 1$ blocks in labeled column vectors, and $W^{(\pm)}_{k_0\rightarrow k_f}$ are the two-band Wilson lines for non-Hermitian systems in corresponding subspaces.
Likewise, $W^{(\pm)}_{k_0\rightarrow k_f}$ can also be numerically calculated as follows:
\begin{eqnarray}
	W^{(\pm)}_{k_0\rightarrow k}&=&U_{\pm,k}^{(l)\dag} {U}_{\pm,k_{l-1}}^{(r)}\dots {U}_{\pm,k_i}^{(l)\dag} {U}_{\pm,{k}_{i-1}}^{(r)} \dots {U}_{\pm,{k}_1}^{(l)\dag} {U}_{\pm,{k}_0}^{(r)} 
	\notag\\
	&=&U_{\pm,k}^{(l)\dag} P_{\pm,k_{l-1}}\dots P_{\pm,{k}_{i}} \dots P_{\pm,{k}_1} U_{\pm,{k}_0}^{(r)},
\end{eqnarray}
where $P_{\pm,{k}_{i}}\equiv U_{\pm,{k}_{i}}^{(r)} U_{\pm,k_i}^{(l)\dag}$ is the projection matrix to the lowest/highest two bands at $k_i$. Apparently, the Wilson loop (i.e., $k=k_0$) is immune to the local similarity transformations in the subspace due to the invariance of the projection matrices to the transformations \cite{Xu-Zhu-2022}.
Note that the phase accumulation of the Wilson loop $W^{(-)}_{k_0\rightarrow k_0+2\pi}$ is just the Zak phase defined in Eq. \eqref{aeq:Zak}, i.e., $\varphi_\text{Zak}= i\ln \det W^{(-)}_{k_0\rightarrow k_0+2\pi}$.

\section{Non-Abelian dynamics in $k$ space for other initial states}
\label{seca:initial_states}

\begin{figure*}[tb]
	\includegraphics[width=0.8\linewidth]{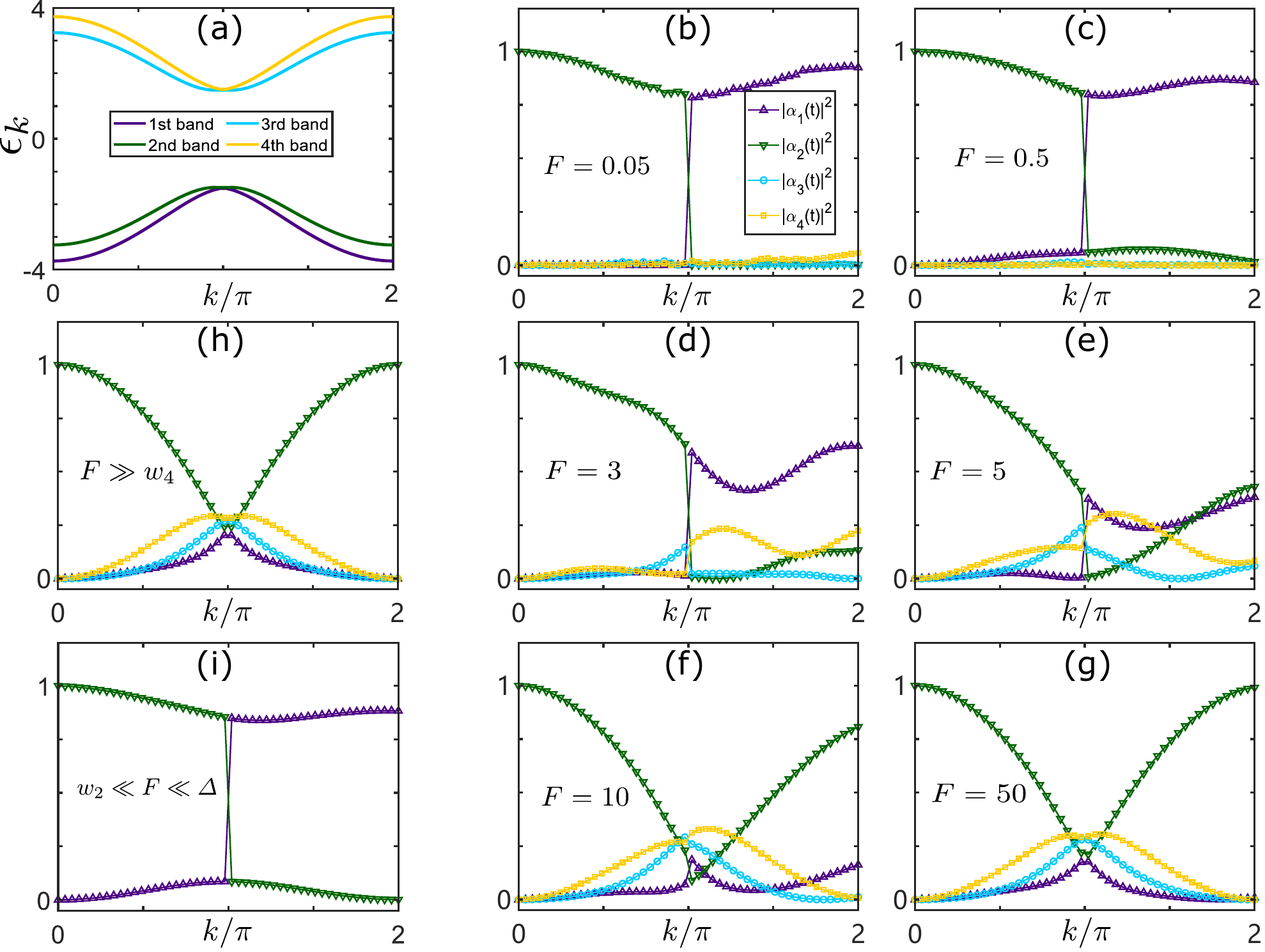}
\caption{
	(a) The identical to Fig. \ref{fig1}(a) of the main text, for convenience to analyze the subsequent figures. 
	(b)--(i) The same settings as in Figs. \ref{fig1}(b)--\ref{fig1}(i) of the main text, except for the initial state being instead located in the second-lowest band at $k=0$ in (a).}
\label{fig4}
\end{figure*}

\begin{figure*}[tb]
	\includegraphics[width=0.8\linewidth]{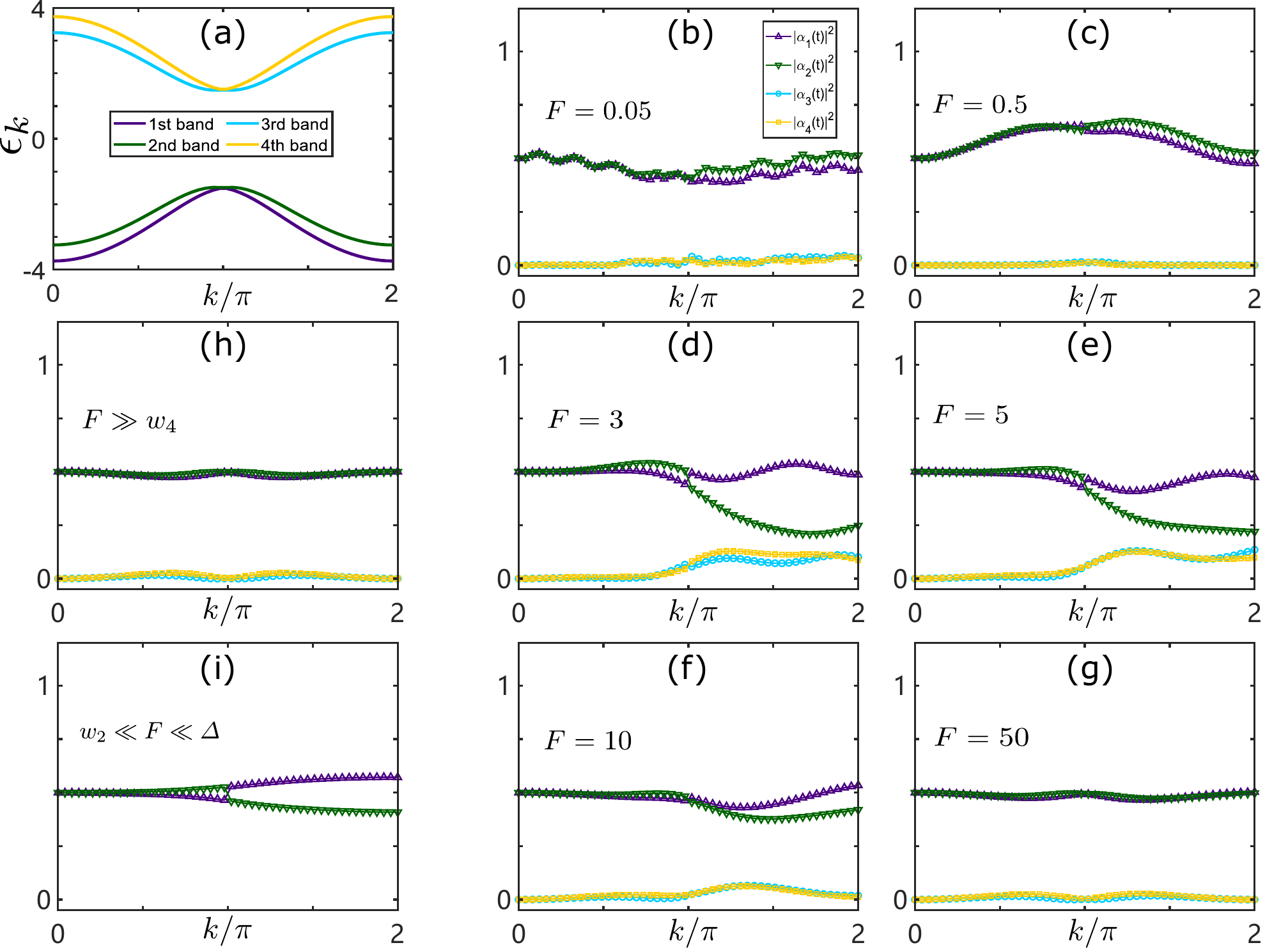}
	\caption{
	(a) The identical to Fig. \ref{fig1}(a) of the main text, for convenience to analyze the subsequent figures. 
	(b)--(i) The same settings as in Figs. \ref{fig1}(b)--\ref{fig1}(i) of the main text, except for the initial state being instead the equal superposition of Bloch states in the lowest two bands at $k=0$ in (a).}
	\label{fig5}
\end{figure*}

In this appendix, we demonstrate the non-Abelian dynamics $|\psi(t)\rangle =  \hat u_{k(t)}^{(r)\dag} \alpha(t) |0\rangle$ in $k$ space in the same way as in Fig. \ref{fig3} of the main text, but for two other initial states: 
(1) the Bloch state located in the second-lowest band at $k=0$, i.e., $\alpha(0)=(0,1,0,0)^T$, and (2) the equal superposition of Bloch states in the lowest two bands at $k=0$, i.e., $\alpha(0)=(1,1,0,0)^T/\sqrt{2}$.
The corresponding figures are plotted in Figs. \ref{fig4} and \ref{fig5}, respectively.

\bibliography{ref}

\end{document}